\documentclass[showpacs,amsmath,amsfonts,amssymb,aps,superscriptaddress]{revtex4}

\usepackage{tabularx}
\usepackage{comment}
\usepackage{enumitem}
\usepackage{graphicx}
\usepackage{float}
\usepackage{dcolumn}
\usepackage{bm}
\usepackage{mathrsfs}
\usepackage{amsmath}

\usepackage{amsfonts}
\usepackage{dsfont}

\begin{document}

\title{Quasinormal modes of noncommutative geometry-inspired dirty black holes}
\author{Davide Batic}
\email{davide.batic@ku.ac.ae}
\affiliation{
Mathematics Department, Khalifa University of Science and Technology, PO Box 127788, Abu Dhabi, United Arab Emirates}

\author{Denys Dutykh}
\email{denys.dutykh@ku.ac.ae}
\affiliation{
Mathematics Department, Khalifa University of Science and Technology, PO Box 127788, Abu Dhabi, United Arab Emirates}

\author{Zeinabou Ahmed Babou}
\email{100060187@ku.ac.ae}
\affiliation{
Mathematics Department, Khalifa University of Science and Technology, PO Box 127788, Abu Dhabi, United Arab Emirates}
\date{\today}

\begin{abstract}
We investigate the quasinormal modes (QNMs) of noncommutative geometry-inspired dirty black holes, focusing on both non-extremal and extremal configurations. These gravitational objects, characterized by smeared energy distributions within a modified de Sitter-like equation of state, modify the classical Schwarzschild metric and regularize central singularities. We employ a spectral method based on Chebyshev polynomials to solve the eigenvalue problem for scalar, electromagnetic, and gravitational perturbations. Our results reveal new overdamped modes indicative of rapid decay without oscillation, particularly prominent in near-extremal and extremal regimes. Additionally, we establish that the QNMs converge to classical Schwarzschild values for large mass parameters, validating our method's robustness. Our findings highlight the impact of dirtiness and noncommutative effects on black hole QNM spectra, offering potential observational signatures for distinguishing these objects in gravitational-wave detections.
\end{abstract}
\maketitle

\section{Introduction}

Black holes have long served as natural laboratories to probe the interplay between gravity and quantum theory. One of their most striking features is represented by the quasinormal mode spectrum, which governs how a perturbed black hole returns to equilibrium and encodes information about the underlying spacetime geometry. Over the years, various approaches ranging from the continued fraction method to Wentzel-Kramers-Brillouin (WKB) approximations and, more recently, spectral methods have been employed to explore QNMs in diverse black hole backgrounds \cite{Leaver1985PRSLA, IYER1987PRD, Mamani2022EPJC, Batic2024EPJC, Batic2024PRD, Konoplya2003ATMP, Konoplya2010PRD, Konoplya2016JCAP, Konoplya2005PRD, Konoplya2011RMP, Konoplya2018PLB, Konoplya2019PRD, Konoplya2022PRL}.

Noncommutative geometry has drawn significant attention as a potential framework for addressing the longstanding singularity issues in general relativity and incorporating quantum corrections at short distances. While a definitive theory of quantum gravity is not yet available, existing candidates such as string theory, loop quantum gravity, and noncommutative geometry exhibit several shared features. These include the noncommutativity of spacetime coordinates at a characteristic length scale $\sqrt{\theta}$ \cite{Nicolini2009IJMPA, Madore2000, Chamseddine1993CMP, Connes1995JMP}, a generalized uncertainty principle that accounts for gravitational effects \cite{Adler1999MPLA, Veneziano1986EL, Bambi2008CQG}, the resolution of physical singularities \cite{Parker1973PR, Sami2006PRD, Kreienbuehl2009PRD}, and the prediction of black hole remnants \cite{Giddings1992PRD, Li2007PLB}.

The term {\it{dirty}} or {\it{hairy}} black holes typically refers to solutions of the Einstein field equations that involve interactions with various types of matter fields. For instance, the coupling of gravity with electromagnetism and a dilaton has been studied in \cite{Gibbons1988NPB, Ichinose1989MPLA, Yamazaki1992CQG, Garfinkle1991PRD, Garfinkle1992PRD}. The interaction of gravity with electromagnetism and an axion was explored in \cite{Allen1990PLB, Campbell1991PLB, Lee1991PRD}, while the addition of Abelian Higgs fields to gravity and electromagnetism was investigated in \cite{Dowker1992PRD}. The combination of gravity, electromagnetism, dilaton, and axion was examined in \cite{Shapere1991MPLA}, and gravity coupled to non-Abelian gauge fields was addressed in \cite{Galtsov1989PLA, Straumann1990PLB, Bizon1990PRL, Bizon1991PLB}. Finally, the interaction of gravity with axion and non-Abelian gauge fields was analyzed in \cite{Lahiri1992PLB}. In the framework of noncommutative geometry-inspired models \cite{Nicolini2009IJMPA}, a dirty black hole is instead characterized by the introduction of a smeared energy and pressure distribution rather than explicit coupling to external matter fields. This smeared distribution serves as a classical parameterization for the energy and pressure stored in an underlying field. Additionally, a de Sitter-like equation of state is assumed to relate the matter density to the radial pressure (see equation (2.14) in \cite{Nicolini2010CQG}).   This approach leads to modifications that regularize the central singularity \cite{Nicolini2006PLB}. Furthermore, for specific ranges of the mass parameter, the gravitational object may transition into a horizonless minigravastar \cite{Nicolini2010CQG}.

It is interesting to observe that if one assumes the fundamental noncommutative scale to be of the order of the Planck length, $\ell_p \approx 1.6 \times 10^{-35}$ m, then one might expect any observable noncommutative effects to be relevant only for black holes of extremely small mass, which would evaporate almost instantaneously. However, as emphasized in \cite{Nicolini2009IJMPA} and further supported by a number of studies in the literature, noncommutative geometry is expected to be significant over a window $\ell_p < \sqrt{\theta} < 10^{-16}$ m. There are theoretical studies that have explored possible constraints on the noncommutative parameter based on indirect experimental evidence or theoretical considerations \cite{AmelinoCamelia2011PRD, Girelli2007PRD}. For instance, studies addressing noncommutative  Quantum Electrodynamics effects have yielded a noncommutative parameter in the range $6~\mbox{GeV}\div 1.7~\mbox{TeV}$ \cite{Abbiendi2003PLB,Hewett2001PRD,Stern2008PRL}. Using the conversion factor $\lambda=hc/E$ where $\lambda$ is the wavelength, $h$ is Planck's constant, $c$ the speed of light, and $E$ the energy, the corresponding range for $\sqrt{\theta}$ is $7.3\cdot 10^{-19}\div 2\cdot 10^{-16}$ m. Similarly, quantum gravity considerations have led to an upper bound of $\sqrt{\theta}\lesssim 9.51\times 10^{-18}$ m \cite{Saha2007EPJC}. On the other hand, by means of the theoretical limit of the Lamb shift in the hydrogen atom \cite{Stern2008PRD} found that $\sqrt{\theta}\lesssim 2.1\cdot 10^{-15}$ m. Moreover, from Cosmic Microwave Background (CMB) radiation data \cite{Akofor2009PRD} obtained $\sqrt{\theta}\leq 1.2\cdot 10^{-19}$ m while bounds emerging from particle phenomenology \cite{Joseph2009PRD} indicate that $\sqrt{\theta}\leq 10^{-31}$ m.  While a thorough examination of bounds on the noncommutative parameter can be found in \cite{Szabo2010GRG}, it may not reflect the latest findings. Therefore, we would like to highlight some more recent research. Notably, in \cite{Moumni2011JGP}, the spectrum of the hydrogen atom was analyzed using noncommutative quantum mechanics, which led to the unsharp constraint of $\sqrt{\theta}\lesssim 4.1\cdot 10^{-12}$ m. Several studies have explored the effects of noncommutativity in quantum systems. For instance, by investigating the Pauli oscillator in noncommutative space, \cite{Heddar2021MPLA} found that $\sqrt{\theta}<1.6\cdot 10^{-13}$ m. \cite{Gnatenko2018IJMPA} studied noncommutative phase space for composite systems, focusing on the impact of noncommutativity on exotic atom spectra, and obtained an upper bound of $\sqrt{\theta}\leq 10^{-14}$ m, which was further improved to $\sqrt{\theta}\leq 10^{-18}$ m by \cite{Gnatenko2017IJMPA} and confirmed in \cite{Gnatenko2014IJMPA}. Following a different strategy, \cite{Rodriguez2018PRA} analyzed quantum effects of Aharonov-Bohm type in noncommutative space and found a limit of $\sqrt{\theta}\approx 9.54\cdot 10^{-18}$ m, improving previous estimates obtained by \cite{Chaichian2002PLB}. Other upper bounds on noncommutative spacetime have been derived from the CMB data of the PLANCK space mission, yielding $\sqrt{\theta}\lesssim 10^{-19}$ m \cite{Joby2015PRD}, and from the study of the relativistic spectrum of the hydrogen atom with spacetime noncommutativity, leading to a constraint of the same order \cite{Moumni2011AIPCP}. Finally, an alternative approach to probe NG via quantum optical experiments has been proposed by \cite{Dey2017NPB}.

This broader window implies that noncommutative corrections may be observable in astrophysical scenarios where the effective noncommutative parameter is much larger than $\ell_p$. In such cases, the resulting QNM spectra could carry signatures of these quantum gravitational corrections without the need to invoke transient, sub-Planckian black holes.

Motivated by the lack of previous studies, to the best of our knowledge, on the QNMs for noncommutative geometry-inspired dirty black holes, we undertake a detailed investigation of such modes here. Our analysis focuses on both non-extremal and extremal configurations, and we compare our findings with the corresponding QNMs of the noncommutative geometry-inspired Schwarzschild black hole \cite{Nicolini2006PLB, Batic2024EPJC}.

QNMs of black holes have been studied using a variety of techniques, each offering unique advantages and limitations. Among the most established methods is the WKB approach, which draws an analogy with quantum mechanical tunnelling problems and the Schr\"{o}dinger equation. This method approximates quasinormal frequencies near the maximum of the effective potential but can face convergence issues, particularly when the second derivative of the potential vanishes at its maximum or when higher precision is needed \cite{IYER1987PRD, Konoplya2003ATMP, Konoplya2010PRD, Konoplya2018PLB, Konoplya2022PRL}.  Another widely used technique is the inverted potential method, which is grounded in group theory and primarily applicable to problems where effective potential governing field perturbations can be nicely approximated by a reversed P\"{o}schl–Teller potential \cite{Ferrari1984PRD}. The continued fractions method, introduced by \cite{Leaver1985PRSLA, Leaver1986PRD}, has become a gold standard for numerical accuracy in QNM studies. By expanding the solution into a series with coefficients satisfying a recurrence relation, this method allows for highly precise calculations of QNMs. Despite its robustness, it can become cumbersome for analyzing complex spacetimes characterized by metric coefficients expressed in terms of special functions. The Laplace transform method, as employed by \cite{Nollert1992PRD}, reformulates the radial perturbation equation into an integral transform in the frequency domain, enabling the extraction of quasinormal frequencies. This approach is particularly effective for solving time-domain equations but may require specialized techniques to handle certain boundary conditions.  Another promising method is the asymptotic iterative method (AIM), originally proposed in \cite{Ciftci2003JPA} and adapted for various black hole configurations such as Schwarzschild, Reissner–Nordstr\"{o}m, noncommutative geometry-inspired Schwarzschild, and Kerr spacetimes \cite{Batic2019EPJC}. AIM uses an iterative procedure that can be highly sensitive to initial parameter choices. For a detailed exploration of this topic, we direct the reader to \cite{Batic2023MMAS}. While improper initialization can lead to instability or require many iterations, this method has proven versatile and effective when carefully implemented. More recently, the spectral method has emerged as a powerful alternative, as demonstrated in \cite{Mamani2022EPJC, Batic2024CGG, Batic2024EPJC, Batic2024PRD}. This technique uses expansions in Tchebyshev polynomials to represent solutions across the entire domain. Unlike WKB or inverted potential methods, it does not rely on local approximations of the potential or specific assumptions about angular momentum. Moreover, it achieves fast convergence by appropriately accounting for the singular behaviour of the solution, making it particularly well-suited for handling complex geometries and boundary conditions. Clearly, each method has its strengths and limitations. Local approximation techniques, such as WKB and inverted potential methods, are conceptually straightforward but can struggle with complex spacetimes. High-accuracy methods like continued fractions and Laplace transforms are computationally intensive for intricate geometries. AIM offers flexibility but demands careful tuning to avoid stability issues. While less established in QNM studies, the spectral method combines numerical precision with a global representation of the solution, providing a robust framework for investigating black hole perturbations without requiring heavy approximations.

To compute the QNMs of noncommutative geometry-inspired dirty black holes, we adopt a spectral method similar to that employed in \cite{Mamani2022EPJC, Batic2024EPJC, Batic2024CGG}. Beginning with the perturbation equations for massless scalar, electromagnetic, and gravitational fields, we reformulate them as an eigenvalue problem over a compact domain. This framework uses Chebyshev polynomials as basis functions, enabling fast convergence and high precision in the frequency domain.  Our analysis reveals the presence of purely imaginary QNMs, i.e. overdamped modes indicative of a rapid return to equilibrium without oscillation, in both extremal and near-extremal cases. Furthermore, we validate the robustness of our method by demonstrating that, for large values of the rescaled mass parameter, the QNMs of noncommutative geometry-inspired dirty black holes converge to those of the classic Schwarzschild black hole. This serves as a key benchmark, reinforcing the accuracy and reliability of our numerical approach.

This paper is structured as follows. Section II provides an overview of the equations of motion and the corresponding metric functions for both non-extremal and extremal regimes, discussing the key properties of the line element. Section III details the spectral method employed to compute the QNMs and describes the numerical implementation of our approach. Section IV presents our results, including an in-depth discussion of the appearance of overdamped modes. Finally, Section V offers concluding remarks and explores possible directions for future research.

\section{Equations of motion}

\begin{figure}
\includegraphics[scale=0.35]{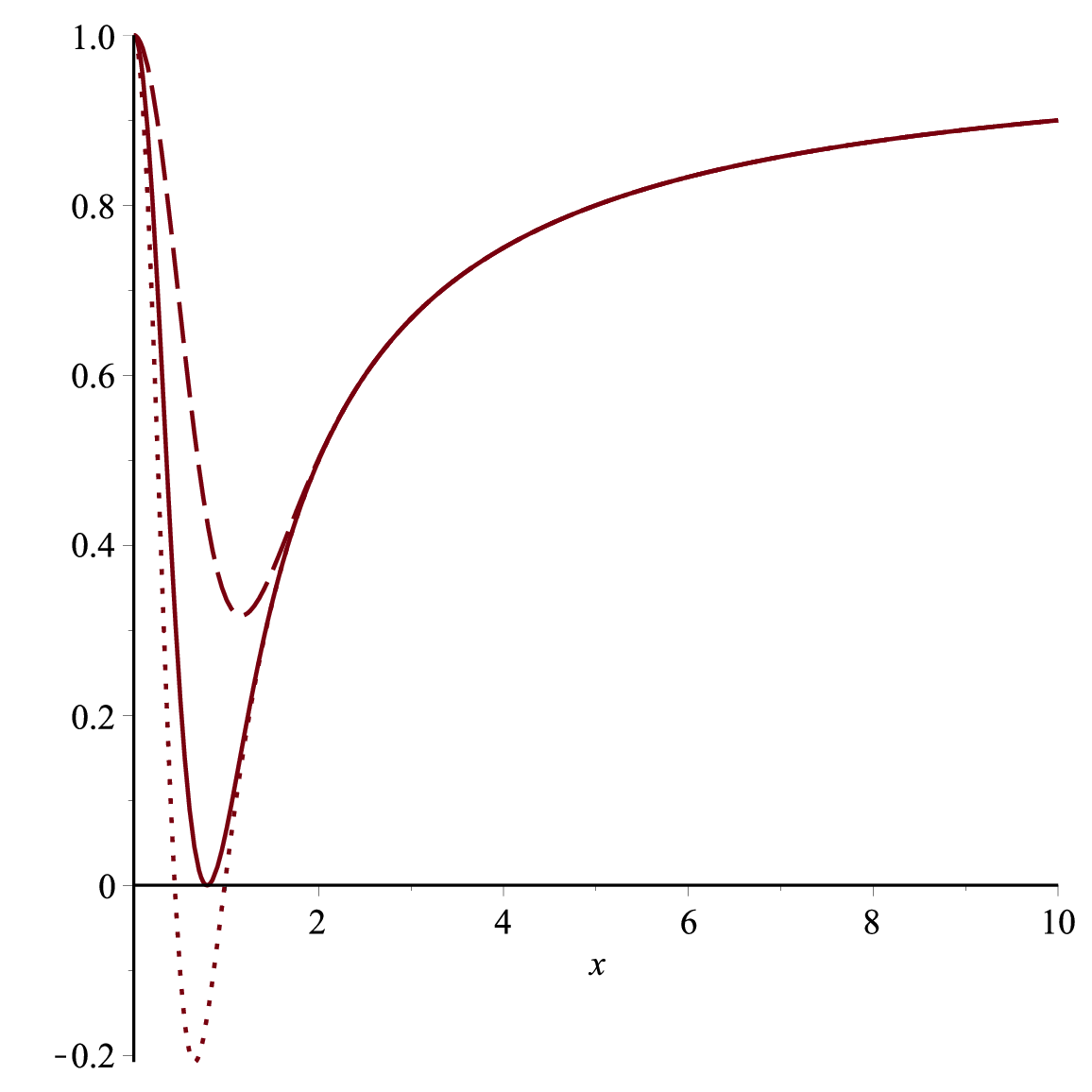}
\caption{\label{fig0}
The plot of the metric coefficient $g^{rr} = f(x)$ as defined in \eqref{f}. An extremal dirty black hole appears when $\mu_e = 1.904119076\ldots$ (solid line), with the horizon located at $x_e = 0.7936575898\ldots$. For $\mu > \mu_e$, a non-extremal dirty black hole with two horizons is present (illustrated by the dotted line for $\mu = 2.3$). When $0 < \mu < \mu_e$, a naked minigravastar emerges (represented by the dashed line for $\mu = 1.3$).}
\end{figure}

We consider a massless scalar field $\psi$ in the background of a noncommutative geometry-inspired dirty black hole. The line element, in units where $c = G_N = 1$, is given by \cite{Nicolini2010CQG}
\begin{eqnarray}
 ds^2 &=& -f(r)e^{-g(r)} dt^2 + \frac{dr^2}{f(r)} + r^2 d\vartheta^2 + r^2 \sin^2{\vartheta} d\varphi^2, \label{metric} \\
 f(r) &=& 1 - \frac{2m(r)}{r},\quad
 g(r)=\frac{M}{2\sqrt{\theta}}\left[1 -\frac{2}{\sqrt{\pi}}\gamma\left(\frac{3}{2}, \frac{r^2}{4\theta}\right)\right]
\end{eqnarray}
with $\vartheta \in [0, \pi]$ and $\varphi \in [0, 2\pi[$. The mass function is defined by
\begin{equation}
 m(r) = \frac{2M}{\sqrt{\pi}} \gamma\left(\frac{3}{2}, \frac{r^2}{4\theta}\right), \qquad \gamma\left(\frac{3}{2}, \frac{r^2}{4\theta}\right) = \int_0^{r^2/4\theta} dt \, \sqrt{t} e^{-t},
\end{equation}
where $M$ is the total mass of the gravitational object, and $\theta$ is a parameter encoding noncommutativity with the dimension of length squared. Here, $\gamma(\cdot, \cdot)$ denotes the lower incomplete gamma function. Using the relationship between the upper and lower incomplete gamma functions as given by \cite{Abramowitz1972}, we can express the $g^{rr}$ metric coefficient as a combination of the standard Schwarzschild term and a perturbation due to noncommutativity
\begin{equation}\label{fr}
  f(r) = 1 - \frac{2M}{r} + \frac{4M}{\sqrt{\pi} r} \, \Gamma\left(\frac{3}{2}, \frac{r^2}{4\theta}\right).
\end{equation}
As $r/\sqrt{\theta} \to +\infty$, both the term $\Gamma(3/2, r^2/4\theta)$ and the argument of the exponential in \eqref{metric} vanish, recovering the classic Schwarzschild metric. If we introduce the rescaling $x=r/2M$ and $\mu=M/\sqrt{\theta}$ together with the identities of the lower incomplete Gamma function \cite{Abramowitz1972}, we can rewrite \eqref{fr} as
\begin{equation}\label{f}
  f(x) = 1 - \frac{\mathrm{erf}(\mu x)}{x} + \frac{2\mu}{\sqrt{\pi}} e^{-\mu^2 x^2},
\end{equation}
where \(\mathrm{erf}(\cdot)\) denotes the error function. The picture emerging from Figure~\ref{fig0}, where $f$ is plotted against the rescaled radial variable $x$, is as follows: an extremal dirty black hole appears when the rescaled mass reaches the critical value $\mu_e = 1.904119076\ldots$. In this regime, the Cauchy horizon $x_c$ and the event horizon $x_h$ coincide with their shared value given by $x_e = x_c = x_h = 0.7936575898\ldots$. When $\mu < \mu_{e}$, the line element in \eqref{metric} describes a dirty minigravastar with no horizons, remaining regular at $x = 0$ \cite{Nicolini2010CQG}. Finally, two distinct horizons are present for $\mu > \mu_{e}$. Unlike the Schwarzschild metric, probing into shorter distances, specifically $r \ll \sqrt{\theta}$, reveals the absence of a central curvature singularity, which is instead replaced by a regular de Sitter core \cite{Nicolini2010CQG}. In this work, we focus our analysis on the QNMs for both the non-extremal and extremal cases.
To achieve this, it is useful to introduce the rescaling $z = x / x_h$, which maps the event horizon to $z = 1$ in both extremal and non-extremal cases. Consequently, the $ g^{rr}$ metric coefficient, as given in \eqref{f}, becomes
\begin{equation}\label{fz}
    f(z) = 1 - \frac{\mathrm{erf}\left(\mu x_h z\right)}{x_h z} + \frac{2\mu}{\sqrt{\pi}} e^{-\mu^2 x_h^2 z^2}.
\end{equation}
Since our line element \eqref{metric} is a particular instance of the more general metric
\begin{equation}\label{metric2}
  ds^2 = -e^{2\Lambda(r)} dt^2 + \frac{dr^2}{1 - \frac{b(r)}{r}} + r^2 d\vartheta^2 + r^2 \sin^2{\vartheta}\, d\varphi^2,
\end{equation}
with the identifications $b(r)=2m(r)$ and $e^{2\Lambda(r)}=f(r)e^{-g(r)}$, the governing equation for a massless field perturbation in the background of a noncommutative geometry-inspired dirty black hole can be written as
\begin{equation}\label{ODE01}
    F(r) \frac{d}{dr} \left( F(r) \frac{d\psi_{\omega\ell s}}{dr} \right) + \left[\omega^2 - U_{s,\ell}(r)\right] \psi_{\omega\ell s}(r) = 0,
\end{equation}
where the field has a time dependence \(e^{-i\omega t}\) and an angular dependence described by spherical harmonics. Here, the parameter $s$ distinguishes between different types of perturbations: $s=0$ corresponds to a massless scalar field, $s=1$ to electromagnetic perturbations, and $s=2$ to vector-type gravitational perturbations \cite{Batic2024CGG}. The function $F(r)$ is defined as $F(r) = f(r) e^{-g(r)}$, and the effective potential \( U_{s,\ell}(r) \) takes the form
\begin{equation}
U_{s,\ell}(r) = f(r) e^{-2g(r)} \left[\frac{\ell(\ell+1)}{r^2} + \frac{1 - s^2}{r} \frac{df}{dr} - \frac{1 - s}{r} f(r) \frac{dg}{dr}\right],
\end{equation}
where $\ell = 0, 1, 2, \ldots$. Note that when $g = 0$, the effective potential reproduces the corresponding effective potential derived for the noncommutative geometry-inspired Schwarzschild black hole in \cite{Batic2024EPJC}.
Since $r = 2M x_h z$ and $\mu = M / \sqrt{\theta}$, equation \eqref{ODE01} can be rewritten as
\begin{eqnarray}
&& \frac{F(z)}{4} \frac{d}{dz} \left(F(z) \frac{d\psi_{\Omega \ell s}}{dz}\right) + \left[x_h^2 \Omega^2 - V_s(z)\right] \psi_{\Omega \ell s}(z) = 0, \qquad \Omega = M \omega, \label{ourODE}\\
&& V_s(z) = \frac{f(z)}{4} e^{-2g(z)} \left[\frac{\ell(\ell+1)}{z^2} + \frac{1 - s^2}{z} \frac{df}{dz} - \frac{1 - s}{z} f(z) \frac{dg}{dz}\right],
\end{eqnarray}
where $f(z)$ is given by \eqref{fz}, and
\begin{equation}
g(z) = \frac{\mu}{2} \left[1 - \mathrm{erf}\left(\mu x_h z\right) + \frac{2\mu x_h z}{\sqrt{\pi}} e^{-\mu^2 x_h^2 z^2}\right].
\end{equation}
As a consistency check, setting $g = 0$ (corresponding to the noncommutative Schwarzschild case) in \eqref{ourODE} reproduces equation (10) from \cite{Batic2024EPJC}. Our analysis focuses on computing the QNMs for the spectral problem stated in \eqref{ourODE}. To this end, we express $\Omega$ as $\Omega = \Omega_R + i \Omega_I$, where $\Omega_I < 0$ ensures that the perturbation decays over time.  The boundary conditions are chosen so that the radial field exhibits inward radiation at the event horizon and outward radiation at spatial infinity. This requires a detailed examination of the asymptotic behaviour of the solution in \eqref{ourODE}, both near the event horizon ($z \to 1^+$) and at large spatial distances ($z \to +\infty$). Additionally, to compute the QNMs using the spectral method, we recast the differential equation in \eqref{ourODE} and the corresponding boundary conditions over the compact interval $[-1, 1]$. This transformation is crucial, as the spectral method represents the regular part of the eigenfunctions using Chebyshev polynomials.

\subsection{The non-extreme case $\mu > \mu_e$}

The metric coefficient $g_{tt}$ is given as defined in (\ref{metric}). It is important to note that, in this case, the Cauchy and event horizons are distinct, meaning that $g_{tt}$ exhibits a simple zero at $z = 1$. To establish the QNM boundary conditions at the event horizon and at infinity, we first need to determine the asymptotic behaviour of the radial solution $\psi_{\Omega \ell s}$ as $z \to 1^+$ and $z \to +\infty$. From this asymptotic data, we can then extract the QNM boundary conditions.  As a consistency check, we will verify that in the limit $g \to 0$, these conditions correctly reproduce those for the noncommutative geometry-inspired Schwarzschild black hole as specified in equation (28) of \cite{Batic2024EPJC}. We conduct our analysis by examining the behaviour of the radial field in two separate regions. 
\begin{enumerate}
\item 
{\underline{Asymptotic behavior as $z\to 1^+$}}: Given that $z = 1$ is a simple zero of $g_{tt}$, we can express it in the form $f(z)e^{-g(z)} = (z - 1)h(z)e^{-g(z)}$, where $h$ is an analytic function at $z = 1$ and satisfies $h(1) = f'(1) \neq 0$. Here, the prime symbol denotes differentiation with respect to \( z \). This formulation allows us to rewrite equation (\ref{ourODE}) in the following form
\begin{eqnarray}
    &&\frac{d^2\psi_{\Omega \ell s}}{dz^2} + p(z) \frac{d\psi_{\Omega \ell s}}{dz} + q(z)\psi_{\Omega \ell s}(z) = 0, \label{ODEZ} \\
    &&p(z) = \frac{1}{z - 1} + \frac{h'(z)}{h(z)} - g'(z), \\
    &&q(z) = \frac{4x_h^2 \Omega^2 e^{2g(z)}}{(z - 1)^2 h^2(z)} - \frac{1}{(z - 1)h(z)} \left[\frac{\ell(\ell + 1)}{z^2} + \frac{1 - s^2}{z} \frac{df}{dz} - \frac{1 - s}{z} f(z) \frac{dg}{dz}\right].
\end{eqnarray}
Since $p$ and $q$ have poles of order one and two at $z = 1$, respectively, this point qualifies as a regular singular point of (\ref{ODEZ}) by Frobenius theory \cite{Ince1956}. Consequently, we can construct solutions in the form
\begin{equation}
    \psi_{\Omega \ell s}(z) = (z - 1)^\rho \sum_{\kappa = 0}^\infty a_\kappa (z - 1)^\kappa.
\end{equation}
The leading behavior at $z = 1$ is represented by the term $(z - 1)^\rho$, where $\rho$ is determined by the indicial equation
\begin{equation}\label{indicial}
    \rho(\rho - 1) + P_0 \rho + Q_0 = 0,
\end{equation}
with
\begin{equation}
    P_0 = \lim_{z \to 1} (z - 1)p(z) = 1, \qquad
    Q_0 = \lim_{z \to 1} (z - 1)^2 q(z) = \left(\frac{2x_h \Omega e^{g(1)}}{f'(1)}\right)^2.
\end{equation}
The roots of (\ref{indicial}) are $\rho_\pm = \pm 2i x_h \Omega e^{g(1)}/f'(1)$, and the appropriate QNM boundary condition at $z = 1$ is
\begin{equation}\label{QNMBCz1}
    \psi_{\Omega \ell s} \underset{{z \to 1^+}}{\longrightarrow} (z - 1)^{-2i x_h \alpha \Omega}, \quad \alpha = \frac{e^{g(1)}}{f'(1)}.
\end{equation}
A compact formula for $f'(1)$ and $e^{g(1)}$ can be obtained by observing that 
\begin{equation}\label{fp1}
    f'(1) = \frac{\mathrm{erf}(\mu x_h)}{x_h} - \frac{2\mu}{\sqrt{\pi}}(1 + 2\mu^2 x_h^2)e^{-\mu^2 x_h^2}
\end{equation}
can be further simplified by using the condition $f(1) = 0$. Solving the latter with respect to the error function yields
\begin{equation}
    \mathrm{erf}(\mu x_h) = x_h \left(1 + \frac{2\mu}{\sqrt{\pi}} e^{-\mu^2 x_h^2}\right),
\end{equation}
which, in turn, allows us to express (\ref{QNMBCz1}) as
\begin{equation}\label{alphanonext}
    \alpha = \frac{e^{-\frac{\mu}{2}(x_h - 1)}}{1 - \frac{4\mu^3 x_h^2}{\sqrt{\pi}} e^{-\mu^2 x_h^2}}.
\end{equation}
A remark is now in order. Setting $g = 0$ in (\ref{QNMBCz1}) correctly reproduces the corresponding QNM boundary condition at the event horizon for a noncommutative geometry-inspired Schwarzschild black hole, as found in equation (16) of \cite{Batic2024EPJC}.
\item
{\underline{Asymptotic behaviour as $z\to+\infty$}}: We begin by rewriting equation (\ref{ourODE}) as
\begin{equation}\label{ODEas}
    \frac{d^2\psi_{\Omega \ell \epsilon}}{dz^2} + P(z) \frac{d\psi_{\Omega \ell \epsilon}}{dz} + Q(z) \psi_{\Omega \ell \epsilon}(z) = 0, \qquad
    P(z) = \frac{f'(z)}{f(z)} - g'(z), \qquad
    Q(z) = \frac{4\left[x_h^2 \Omega^2 - V_s(z)\right]}{f^2(z)e^{-2g(z)}},
\end{equation}
where $V_s(z)$ is defined in (\ref{ourODE}). The asymptotic behaviour of solutions to (\ref{ODEas}) can be effectively deduced using the method described in \cite{Olver1994MAA}. To proceed, we observe that
\begin{equation}
    P(z) = \sum_{\kappa=0}^\infty \frac{\mathfrak{f}_\kappa}{z^k} = \mathcal{O}\left(\frac{1}{z^2}\right), \qquad
    Q(z) = \sum_{\kappa=0}^\infty \frac{\mathfrak{g}_\kappa}{z^k} = 4x_h^2 \Omega^2 + \frac{8x_h \Omega^2}{z} + \mathcal{O}\left(\frac{1}{z^2}\right).
\end{equation}
Given that at least one of the coefficients \( \mathfrak{f}_0 \), \( \mathfrak{g}_0 \), or \( \mathfrak{g}_1 \) is non-zero, a formal solution to (\ref{ODEas}) is given by \cite{Olver1994MAA}
\begin{equation}\label{olvers}
    \psi^{(j)}_{\Omega \ell s}(z) = z^{\mu_j} e^{\lambda_j z} \sum_{\kappa=0}^\infty \frac{a_{\kappa, j}}{z^\kappa}, \qquad j \in \{1,2\},
\end{equation}
where \( \lambda_1 \), \( \lambda_2 \), \( \mu_1 \), and \( \mu_2 \) are the roots of the characteristic equations
\begin{equation}\label{chareqns}
    \lambda^2 + \mathfrak{f}_0 \lambda + \mathfrak{g}_0 = 0, \quad
    \mu_j = -\frac{\mathfrak{f}_1 \lambda_j + \mathfrak{g}_1}{\mathfrak{f}_0 + 2 \lambda_j}.
\end{equation}
A straightforward calculation shows that \( \lambda_\pm = \pm 2i x_h \Omega \) and \( \mu_\pm = \pm 2i \Omega \). Consequently, the QNM boundary condition at spatial infinity can be expressed as
\begin{equation}\label{QNMBCzinf}
    \psi_{\Omega \ell s} \underset{{z \to +\infty}}{\longrightarrow} z^{2i \Omega} e^{2i x_h \Omega z}.
\end{equation}
Observing that this result accurately reproduces the corresponding boundary condition for the noncommutative geometry-inspired Schwarzschild case is satisfying.
\end{enumerate}
At this stage, we can proceed as in \cite{Batic2024EPJC} by transforming the radial function $\psi_{\Omega \ell s}(z)$ into a new radial function $\Phi_{\Omega \ell s}(z)$ such that the QNM boundary conditions are automatically satisfied. In this form, $\Phi_{\Omega \ell s}(z)$ remains regular at both $z = 1$ and spatial infinity. To achieve this, we first apply the transformation
\begin{equation}\label{Ansatz}
    \psi_{\Omega \ell s}(z) = z^{2i(1 + x_h \alpha)\Omega} (z - 1)^{-2i x_h \alpha \Omega} e^{2i x_h \Omega (z - 1)} \Phi_{\Omega \ell s}(z),
\end{equation}
where  $\alpha$ is defined by \eqref{alphanonext}, followed by the change of variable $z=2/(1-y)$, which maps the point at infinity and the event horizon to $y = 1$ and $y = -1$, respectively. This leads to the differential equation
\begin{equation}\label{ODEynone}
    S_2(y)\ddot{\Phi}_{\Omega\ell s}(y) + S_1(y)\dot{\Phi}_{\Omega\ell s}(y) + S_0(y)\Phi_{\Omega\ell s}(y) = 0,
\end{equation}
where a dot represents differentiation with respect to the variable $y$ and 
\begin{eqnarray}
  S_2(y) &=& \frac{(1+y)^2}{4} F^2(y), \label{S2onone} \\
  S_1(y) &=& i\Omega\frac{1+y}{(1-y)^2}F^2(y)\left[(1+y)(1+2x_h-y)-x_h\alpha(1-y)^2\right]-\frac{(1+y)^2}{2(1-y)}F^2(y)+\frac{(1+y)^2}{4} F(y)\dot{F}(y), \label{S1onone}\\
  S_0(y) &=& \Omega^2\Sigma_2(y)+i\Omega\Sigma_1(y)+\Sigma_0(y) \label{S0onone}
\end{eqnarray}
with
\begin{eqnarray}
    \Sigma_2(y) &=& \frac{4x_h^2(1+y)^2}{(1-y)^4}-\frac{F^2(y)}{(1-y)^4}\left[(1+y)(1+2x_h-y)-x_h\alpha(1-y)^2\right]^2,\\
    \Sigma_1(y) &=& \frac{F(y)}{2}\left\{
    \left(\frac{1+y}{1-y}\right)^2\left[(1+2x_h-y)\dot{F}(y)-F(y)\right]+\frac{x_h\alpha}{1-y}\left[(3+y)F(y)-(1-y^2)\dot{F}(y)\right]
    \right\},\\
    \Sigma_0(y) &=& -\frac{4(1+y)^2}{(1-y)^4}V_s(y).
\end{eqnarray}
In addition, we also require that $\Phi_{\Omega\ell s}(y)$ remains regular at $y=\pm 1$. As a consequence of the transformation introduced above, we have
\begin{eqnarray}
F(y)&=&f(y)e^{-g(y)},\label{fv}\\
f(y)&=& 1 - \frac{1-y}{2x_h}\mbox{erf}\left(\frac{2\mu x_h}{1-y}\right) + \frac{2\mu}{\sqrt{\pi}}e^{-\frac{4\mu^2 x_h^2}{(1-y)^2}},\quad
g(y)=\frac{\mu}{2}\left[1 -\mbox{erf}\left(\frac{2\mu x_h}{1-y}\right)+\frac{4\mu x_h}{\sqrt{\pi}(1-y)}e^{-\frac{4\mu^2 x_h^2}{(1-y)^2}}\right],\label{fv0}\\
V_s(y) &=& \frac{(1-y)^2}{16}f(y)e^{-2g(y)}\left[(1-s^2)(1-y)\dot{f}(y)-(1-s)(1-y)f(y)\dot{g}(y)+\ell(\ell+1)\right].\label{fv1}
\end{eqnarray}
\begin{table}
\caption{Classification of the points $y=\pm 1$ for the relevant functions defined by  (\ref{S2onone})-(\ref{S0onone}), and (\ref{fv})-(\ref{fv1}). The abbreviations $z$ ord $n$ and $p$ ord $m$ stand for zero of order $n$ and pole of order $m$, respectively.}
\begin{center}
\begin{tabular}{ | c | c | c | c | c | c | c | c }
\hline
$y$  & $F(y)$  & $V_s(y)$ & $S_2(y)$ & $S_1(y)$ & $S_0(y)$\\ \hline
$-1$ & z \mbox{ord} 1 & z \mbox{ord} 1 & z \mbox{ord} 4& z \mbox{ord} 3 & z \mbox{ord} 3 \\ \hline
$+1$ & $+1$  & z \mbox{ord} 2 & $+1$ & p \mbox{ord} 2 & p \mbox{ord} 2\\ \hline
\end{tabular}
\label{tableEinsnone}
\end{center}
\end{table}
Notice that the coefficients of the differential equation (\ref{ODEynone}) exhibit a common zero of order $3$ at $y = -1$ while $y = 1$ is a pole of order $2$ for the coefficients $S_1$ and $S_0$ (see Table~\ref{tableEinsnone}). This means that in order to apply the spectral method, it is necessary to multiply (\ref{ODEynone}) by $(1-y)^2/(1+y)^3$. Hence, we end up with the following differential equation
\begin{equation}\label{ODEhynone}
    M_2(y)\ddot{\Phi}_{\Omega\ell s}(y) + M_1(y)\dot{\Phi}_{\Omega\ell s}(y) + M_0(y)\Phi_{\Omega\ell s}(y) = 0,
\end{equation}
where
\begin{equation}\label{S210honone}
  M_2(y) = \frac{(1-y)^2}{4(1+y)}F^2(y), \qquad
  M_1(y) = i\Omega N_1(y)+N_0(y), \qquad
  M_0(y) = \Omega^2 C_2(y)+i\Omega C_1(y)+C_0(y)
\end{equation}
with
\begin{eqnarray}
    N_1(y) &=& F^2(y)\left[\frac{1+2x_h-y}{1+y}-x_h\alpha\left(\frac{1-y}{1+y}\right)^2\right], \quad
    N_0(y) = \frac{F(y)}{4(1+y)}\frac{d}{dy}\left((1-y)^2 F(y)\right),\label{N0}\\
    C_2(y) &=& \frac{4x_h^2}{(1+y)(1-y)^2}-\frac{F^2(y)}{(1+y)^3(1-y)^2}\left[(1+y)(1+2x_h-y)-x_h\alpha(1-y)^2\right]^2,\label{C2}\\
    C_1(y) &=& \frac{F(y)}{2(1+y)}\left\{(1+2x_h-y)\dot{F}(y)-F(y)+x_h\alpha\frac{1-y}{(1+y)^2}\left[(3+y)F(y)-(1-y^2)\dot{F}(y)\right]\right\},\label{C1}\\
    C_0(y) &=& -\frac{4V_s(y)}{(1+y)(1-y)^2}.\label{C0}
\end{eqnarray}
It can be easily verified with Maple that
\begin{eqnarray}
    &&\lim_{y\to 1^{-}}M_2(y)=0=\lim_{y\to -1^{+}}M_2(y),\\
    &&\lim_{y\to 1^{-}}M_1(y)=ix_h\Omega,\quad
    \lim_{y\to -1^{+}}M_1(y)=i\Omega\Lambda_1+\Lambda_0,\\
    &&\lim_{y\to 1^{-}}M_0(y)=A_2\Omega^2 +A_0,\quad
     \lim_{y\to -1^{+}}M_0(y)=B_2\Omega^2+i\Omega B_1+B_0,
\end{eqnarray}
where the coefficients entering in the above expressions are
\begin{eqnarray}
\Lambda_1 &=&\frac{x_h e^{-\frac{\mu}{2}(1-x_h)}}{\sqrt{\pi}}\left(4\mu^3 x_h^2 e^{-\mu^2 x_h^2}-\sqrt{\pi}\right),\quad
\Lambda_0 =\frac{e^{-\mu(1-x_h)}}{4\pi}\left(4\mu^3 x_h^2 e^{-\mu^2 x_h^2}-\sqrt{\pi}\right)^2,\\
  A_2 &=& 1-\frac{\sqrt{\pi}x_h^2 e^{\frac{\mu}{2}(1-x_h)}}{4\mu^3 x_h^2 e^{-\mu^2 x_h^2}-\sqrt{\pi}},\quad A_0=-\frac{\ell(\ell+1)}{8},\label{Acoefnone}\\
  B_2 &=& -\frac{x_h^2+x_h}{\sqrt{\pi}}(4\mu^3 x_h^2 e^{-\mu^2 x_h^2}-\sqrt{\pi})e^{-\frac{\mu}{2}(1-x_h)}\nonumber\\
  &&-\frac{2x_h^2\left[4\mu^7 x_h^5
e^{-2\mu^2 x_h^2}-\sqrt{\pi}\mu^3x_h^2(2\mu^2 x_h^2+\mu x_h+2)e^{-\mu^2 x_h^2}+\pi\right]}{\sqrt{\pi}(4\mu^3 x_h^2 e^{-\mu^2 x_h^2}-\sqrt{\pi})},\label{Dcoefnone}\\
  B_1 &=& \frac{\mathfrak{a}_3 e^{-3\mu^2 x_h^2}+\mathfrak{a}_2 e^{-2\mu^2 x_h^2}+\mathfrak{a}_1 e^{-\mu^2 x_h^2}+\mathfrak{a}_0}{4\pi(4\mu^3 x_h^2 e^{-\mu^2 x_h^2}-\sqrt{\pi})},\\
  B_0 &=& \frac{e^{-\mu(1-x_h)}(4\mu^3 x_h^2 e^{-\mu^2 x_h^2}-\sqrt{\pi})}{8\sqrt{\pi}}\left[\ell(\ell+1)-\frac{1-s^2}{\sqrt{\pi}}(4\mu^3 x_h^2 e^{-\mu^2 x_h^2}-\sqrt{\pi})\right]\label{B0},
\end{eqnarray}
and
\begin{eqnarray}
    \mathfrak{a}_3 &=& 32\mu^9 x_h^6 e^{-\frac{\mu}{2}(1-x_h)}\left[2(1+x_h)e^{-\frac{\mu}{2}(1-x_h)}+\mu x_h^2\right],\\
    \mathfrak{a}_2 &=& -16\sqrt{\pi}\mu^6 x_h^4\left[x_h(\mu^2 x_h^2+\mu x_h+1)e^{-\frac{\mu}{2}(1-x_h)}+3(x_h+1)e^{-\mu(1-x_h)}\right],\\
    \mathfrak{a}_1 &=& 2\pi\mu^3 x_h^2\left[x_h(2\mu^2 x_h^2+\mu x_h+6)e^{-\frac{\mu}{2}(1-x_h)}+6(x_h+1)e^{-\mu(1-x_h)}\right],\\ 
    \mathfrak{a}_0 &=&-\pi^{3/2}\left[2x_h e^{-\frac{\mu}{2}(1-x_h)}+(x_h+1)e^{-\mu(1-x_h)}\right].
\end{eqnarray}
As a final step prior to implementing the spectral method, we rewrite the differential equation (\ref{ODEhynone}) in the following form
\begin{equation}\label{TSCH}
  \widehat{L}_0\left[\Phi_{\Omega\ell s}, \dot{\Phi}_{\Omega\ell s}, \ddot{\Phi}_{\Omega\ell s}\right] +  i\widehat{L}_1\left[\Phi_{\Omega\ell s}, \dot{\Phi}_{\Omega\ell s}, \ddot{\Phi}_{\Omega\ell s}\right]\Omega +  \widehat{L}_2\left[\Phi_{\Omega\ell s}, \dot{\Phi}_{\Omega\ell s}, \ddot{\Phi}_{\Omega\ell s}\right]\Omega^2 = 0
\end{equation}
with
\begin{eqnarray}
  \widehat{L}_0\left[\Phi_{\Omega\ell s}, \dot{\Phi}_{\Omega\ell s}, \ddot{\Phi}_{\Omega\ell s}\right] &=& \widehat{L}_{00}(y)\Phi_{\Omega\ell s} + \widehat{L}_{01}(y)\dot{\Phi}_{\Omega\ell s} + \widehat{L}_{02}(y)\ddot{\Phi}_{\Omega\ell s},\label{L0none}\\
  \widehat{L}_1\left[\Phi_{\Omega\ell s}, \dot{\Phi}_{\Omega\ell s}, \ddot{\Phi}_{\Omega\ell s}\right] &=& \widehat{L}_{10}(y)\Phi_{\Omega\ell s} + \widehat{L}_{11}(y)\dot{\Phi}_{\Omega\ell s} + \widehat{L}_{12}(y)\ddot{\Phi}_{\Omega\ell s}, \label{L1none}\\
  \widehat{L}_2\left[\Phi_{\Omega\ell s}, \dot{\Phi}_{\Omega\ell s}, \ddot{\Phi}_{\Omega\ell s}\right] &=& \widehat{L}_{20}(y)\Phi_{\Omega\ell s} + \widehat{L}_{21}(y)\dot{\Phi}_{\Omega\ell s} + \widehat{L}_{22}(y)\ddot{\Phi}_{\Omega\ell s}.\label{L2none}
\end{eqnarray}
Furthermore, Table~\ref{tableZweinone} provides a summary of the $\widehat{L}_{ij}$ terms from (\ref{L0none})--(\ref{L2none}) along with their limiting values at \(y = \pm 1\).
\begin{table}
\caption{We present the definitions of the coefficients \(\widehat{L}_{ij}\) and their behaviors at the endpoints of the interval \(-1 \leq y \leq 1\). The symbols used in the table are defined in (\ref{S210honone})-(\ref{B0}).}
\begin{center}
\begin{tabular}{ | c | c | c | c | c | c | c | c }
\hline
$(i,j)$  & $\displaystyle{\lim_{y\to -1^+}}\widehat{L}_{ij}$  & $\widehat{L}_{ij}$ & $\displaystyle{\lim_{y\to 1^-}}\widehat{L}_{ij}$  \\ \hline
$(0,0)$ &  $B_0$          & $C_0$                  & $A_0$\\ \hline
$(0,1)$ &  $\Lambda_0$    & $N_0$                  & $0$\\ \hline
$(0,2)$ &  $0$            & $M_2$                  & $0$\\ \hline 
$(1,0)$ &  $B_1$          & $C_1$                  & $0$\\ \hline 
$(1,1)$ &  $\Lambda_1$    & $N_1$                  & $x_h$\\ \hline 
$(1,2)$ &  $0$            & $0$                    & $0$\\ \hline 
$(2,0)$ &  $B_2$          & $C_2$                  & $A_2$\\ \hline
$(2,1)$ &  $0$            & $0$                    & $0$\\ \hline
$(2,2)$ &  $0$            & $0$                    & $0$\\ \hline
\end{tabular}
\label{tableZweinone}
\end{center}
\end{table} 

\subsection{The extreme case $\mu = \mu_e$}

Also, in this case, it is convenient to introduce the rescaling $x=r/(2M)$. However, since the Cauchy and the event horizon coincide, \emph{i.e.} $x_e = x_c = x_h = 0.7936575898\ldots$,  it is useful to introduce a second rescaling given by $\xi = x/x_e$. Hence, equation \eqref{ODE01} can be expressed in the following form
\begin{eqnarray}
&& \frac{F_e(\xi)}{4} \frac{d}{d\xi} \left(F_e(\xi) \frac{d\psi_{\Omega \ell s}}{d\xi}\right) + \left[x_e^2 \Omega^2 - V_s(\xi)\right] \psi_{\Omega \ell s}(\xi) = 0, \qquad \Omega = M \omega, \label{ourODEe}\\
&& V_s(\xi) = \frac{f_e(\xi)}{4} e^{-2g(\xi)} \left[\frac{\ell(\ell+1)}{\xi^2} + \frac{1 - s^2}{\xi} \frac{df_e}{d\xi} - \frac{1 - s}{\xi} f_e(\xi) \frac{dg}{d\xi}\right],\quad
 F_e(\xi)=f_e(\xi)e^{-g(\xi)},
\end{eqnarray}
where $f_e(\xi)$ and $g_e(\xi)$ are given by 
\begin{equation}\label{fze}
f_e(\xi)=1-\frac{\mbox{erf}\left(\mu_e x_e\xi\right)}{x_e\xi}+\frac{2\mu_e}{\sqrt{\pi}}e^{-\mu_e^2 x_e^2\xi^2},\quad
g_e(\xi) = \frac{\mu_e}{2} \left[1 - \mathrm{erf}\left(\mu_e x_e \xi\right) + \frac{2\mu_e x_e \xi}{\sqrt{\pi}} e^{-\mu_e^2 x_e^2 \xi^2}\right].
\end{equation}
In this scenario, $f_e(\xi)$ exhibits a zero of order two at $\xi = 1$, and therefore, $f_e(1) = 0 = f^{'}_e(1)$ where the prime denotes differentiation with respect to $\xi$. This observation leads to the following functional relationships
\begin{equation}\label{erf-exp}  
    e^{-\mu_e^2 x_e^2} = \frac{\sqrt{\pi}}{4\mu_e^3 x_e^2}, \quad  
    \mbox{erf}(\mu_e x_e) = \frac{1 + 2\mu_e^2 x_e^2}{2\mu_e^2 x_e},  
\end{equation}  
which play a crucial role in simplifying subsequent calculations. To derive the QNM boundary conditions at both the event horizon and spatial infinity, it is necessary to first determine the asymptotic behaviour of the radial solution \(\psi_{\Omega\ell s}\) as \(\xi \to 1^{+}\) and as \(\xi \to +\infty\). This asymptotic analysis then enables the extraction of the QNM boundary conditions.
\begin{enumerate}
\item 
{\underline{Asymptotic behaviour as $\xi \to 1^+$}}: Considering that \(\xi = 1\) is a double zero of \(f_e(\xi)\), the function can be expressed in the form \(f_e(\xi) = (\xi - 1)^2 h(\xi)\), where \(h(\xi)\) is analytic at \(\xi = 1\) and satisfies \(h(1) = f_e^{\prime\prime}(1)/2 = \mu_e^2 x_e^2 - 1 \approx 1.2837\). This representation allows us to rewrite (\ref{ourODEe}) in the following form  
\begin{eqnarray}  
    &&\frac{d^2\psi_{\Omega\ell s}}{d\xi^2} + \mathfrak{p}(\xi)\frac{d\psi_{\Omega\ell s}}{d\xi} + \mathfrak{q}(\xi)\psi_{\Omega\ell s}(\xi) = 0, \label{ODEZe} \\  
    &&\mathfrak{p}(\xi) = \frac{2}{\xi - 1} + \frac{h^{\prime}(\xi)}{h(\xi)} - g_e^{\prime}(\xi), \\  
    &&\mathfrak{q}(\xi) = \frac{4x_e^2 \Omega^2 e^{2g_e(\xi)}}{(\xi - 1)^4 h^2(\xi)} - \frac{1}{(\xi - 1)^2 h(\xi)} \left[\frac{\ell(\ell + 1)}{\xi^2} + \frac{1 - s^2}{\xi}f^{\prime}_e(\xi) - \frac{1 - s}{\xi} f_e(\xi)g_e^{\prime}(\xi)\right].  
\end{eqnarray}  
The result above demonstrates that \(\mathfrak{q}\) possesses a fourth-order pole at \(\xi = 1\), making this point an irregular singularity. As a result, Frobenius's theory is not applicable in this scenario. However, a similar analysis to that conducted in \cite{Batic2024EPJC} for the extreme case of a noncommutative geometry-inspired Schwarzschild black hole reveals that \(\xi = 1\) is an irregular singular point of rank one. Consequently, the leading behaviour of the solutions to equation (\ref{ODEZe}) near the event horizon can be determined using the method described in \cite{Olver1994MAA}. To achieve this goal, we begin by noting that the transformation \(\tau = (\xi - 1)^{-1}\), which maps the event horizon to infinity and spatial infinity to zero, takes equation (\ref{ODEZe}) into the form  
\begin{equation}
\frac{d^2\psi_{\Omega\ell\epsilon}}{d\tau^2} + \mathfrak{C}(\tau)\frac{d\psi_{\Omega\ell\epsilon}}{d\tau} + \mathfrak{D}(\tau)\psi_{\Omega\ell\epsilon}(\tau) = 0,
\end{equation}
where the coefficients \(\mathfrak{C}(\tau)\) and \(\mathfrak{D}(\tau)\) are given by  
\begin{equation}
\mathfrak{C}(\tau) = \sum_{\kappa=0}^\infty\frac{\mathfrak{c}_\kappa}{\tau^\kappa} = \mathcal{O}\left(\frac{1}{\tau^2}\right), \quad
\mathfrak{D}(\tau) = \sum_{\kappa=0}^\infty\frac{\mathfrak{d}_\kappa}{\tau^\kappa} = \mathfrak{d}_0 + \frac{\mathfrak{d}_1}{\tau} + \mathcal{O}\left(\frac{1}{\tau^2}\right).
\end{equation} 
The leading coefficients \(\mathfrak{d}_0\) and \(\mathfrak{d}_1\) are explicitly given by 
\begin{equation}
\mathfrak{d}_0 = \frac{4x_e^2\Omega^2 e^{-\mu_e(x_e-1)}}{(\mu_e^2 x_e^2 - 1)^2}, \quad \mathfrak{d}_1 = \frac{4x_e^2\Omega^2(4\mu_e^4 x_e^4 - 3\mu_e^3 x_e^3 - 4\mu_e^2 x_e^2 + 3\mu_e x_e - 4)e^{-\mu_e(x_e-1)}}{3(\mu_e^2 x_e^2 - 1)^3}.
\end{equation}
Since at least one of the coefficients \(\mathfrak{c}_0\), \(\mathfrak{d}_0\), or \(\mathfrak{d}_1\) is nonzero, a formal solution to (\ref{ODEZe}) can be expressed as \cite{Olver1994MAA}
\begin{equation}
\psi^{(\pm)}_{\Omega\ell s}(\tau) = \tau^{\mu_\pm} e^{\lambda_\pm \tau} \sum_{\kappa=0}^\infty \frac{\mathfrak{a}_{\kappa,\pm}}{\tau^\kappa},
\end{equation}
where \(\lambda_\pm\) and \(\mu_\pm\) are determined as the roots of the characteristic equations 
\begin{equation}
\lambda_\pm^2 + \mathfrak{d}_0 = 0, \quad  
\mu_\pm = -\frac{\mathfrak{d}_1}{2\lambda_\pm}.    
\end{equation}
A straightforward computation yields the expressions
\begin{equation}\label{lmu}
\lambda_\pm = \pm\frac{2i x_e \Omega e^{-\frac{\mu_e}{2}(x_e - 1)}}{\mu_e^2 x_e^2 - 1}, \quad  
\mu_\pm = \pm\frac{i x_e \Omega (4\mu_e^4 x_e^4 - 3\mu_e^3 x_e^3 - 4\mu_e^2 x_e^2 + 3\mu_e x_e - 4)e^{-\frac{\mu_e}{2}(x_e - 1)}}{3(\mu_e^2 x_e^2 - 1)^2}.
\end{equation}
It is crucial to observe that a radial field displaying purely inward radiation near the event horizon ($\xi \to 1^+$) transforms, under the change of variables $\tau = (\xi - 1)^{-1}$, into a field that radiates outward as $\tau \to +\infty^{-}$. This transformation necessitates choosing the positive sign in the formulas for $\lambda_\pm$ and $\mu_\pm$. Therefore, the correct boundary condition for QNMs at $\xi = 1$ is given by
\begin{equation}\label{QNMBCe1}
\psi_{\Omega\ell s} \underset{{\xi \to 1^+}}{\longrightarrow} (\xi - 1)^{-\mu_+} \exp\left(\frac{\lambda_+}{\xi - 1}\right),
\end{equation}
where $\mu_+$ and $\lambda_+$ are defined as in equation \eqref{lmu}.
\item 
{\underline{Asymptotic behaviour as $\xi \to +\infty$}}: Since the asymptotic analysis at spatial infinity for the extreme case follows the same procedure as in the non-extreme case, we refer the reader to the detailed derivation presented earlier and simply state that the QNM boundary condition at spatial infinity is
\begin{equation}\label{QNMBCzinfe}
    \psi_{\Omega\ell s} \underset{{\xi \to +\infty}}{\longrightarrow} \xi^{2i\Omega} e^{2i x_e\Omega \xi}.
\end{equation}
\end{enumerate}
To ensure that the radial function $\psi_{\Omega\ell s}(\xi)$ automatically satisfies the QNM boundary conditions at $\xi = 1$ and at spatial infinity, we introduce a transformation that factors out the known asymptotic behavior. This leads us to consider the following ansatz
\begin{equation}\label{Ansatze}
    \psi_{\Omega\ell s}(\xi) = \xi^{2i\Omega + \mu_+} (\xi - 1)^{-\mu_+} \exp\left[2i x_e \Omega (\xi - 1) + \frac{\lambda_+}{\xi - 1}\right] \Phi_{\Omega\ell s}(\xi).
\end{equation}
This initial ansatz can be rearranged into a more compact form, namely
\begin{equation}
\psi_{\Omega\ell s}(\xi) = \xi^{2i a \Omega} (\xi - 1)^{-2i (a - 1) \Omega} \exp\left[2i x_e \Omega \eta(\xi)\right] \Phi_{\Omega\ell s}(\xi)
\end{equation}
by introducing the parameter $a$ and the auxiliary function $\eta(\xi)$ defined as 
\begin{equation}
a = 1 + \frac{x_e \left(4\mu_e^4 x_e^4 - 3\mu_e^3 x_e^3 - 4\mu_e^2 x_e^2 + 3\mu_e x_e - 4\right) e^{-\frac{\mu_e}{2}(x_e - 1)}}{6 (\mu_e^2 x_e^2 - 1)^2}, \quad
    \eta(\xi) = \xi - 1 + \frac{e^{-\frac{\mu_e}{2}(x_e - 1)}}{(\mu_e^2 x_e^2 - 1)(\xi - 1)}. 
\end{equation}
Finally, to facilitate numerical treatment and further analysis, we map the radial coordinate $\xi$ onto a compact interval by performing the transformation $\xi =2/(1-y)$. In this new variable $y$, equation \eqref{ODEZe} transforms into
\begin{equation}\label{ODEye}
    S_{2e}(y)\ddot{\Phi}_{\Omega\ell s}(y)+S_{1e}(y)\dot{\Phi}_{\Omega\ell s}(y)+S_{0e}(y)\Phi_{\Omega\ell s}(y)=0,
\end{equation}
where a dot means differentiation with respect to the variable $y$ and 
\begin{eqnarray}
    S_{2e}(y)&=&\frac{(1+y)^2}{4} F_e^2(y),\label{S2oe}\\
    S_{1e}(y)&=&i\Omega\frac{1+y}{1-y}F_e^2(y)\left[x_e(1-y^2)\dot{\eta}(y)+2-a(1-y)\right]-\frac{(1+y)^2}{2(1-y)}F_e^2(y)+\frac{(1+y)^2}{4}F_e(y)\dot{F}_e(y),\label{S1oe}\\
    S_{0e}(y)&=&\Omega^2\Sigma_{2e}(y)+i\Omega\Sigma_{1e}(y)+\Sigma_{0e}(y)\label{S0oe}
\end{eqnarray}
with
\begin{eqnarray}
\Sigma_{2e}(y)&=&\frac{4x_e^2(1+y)^2}{(1-y)^4}-\frac{F_e^2(y)}{(1-y)^2}\left[x_e(1-y^2)\dot{\eta}(y)+2-a(1-y)\right]^2,\\
\Sigma_{1e}(y)&=&\frac{x_e}{2}(1+y)^2 F_e(y)\dot{F}_e(y)\dot{\eta}(y)+\frac{x_e}{2}\frac{(1+y)^2}{1-y}F_e^2(y)\left[(1-y)\ddot{\eta}(y)-2\dot{\eta}(y)\right]+\nonumber\\
&&\frac{1+y}{2(1-y)}\left[2-a(1-y)\right]F_e(y)\dot{F}_e(y)-\frac{F_e^2(y)}{(1-y)^2}\left[2-2a(1-y)+\frac{a}{2}(1-y)^2\right],\\
\Sigma_{1e}(y)&=&-\frac{4(1+y)^2}{(1-y)^4}V_s(y).
\end{eqnarray}
and the requirement that $\Phi_{\Omega\ell s}(y)$ is regular at $y = \pm 1$. Consequently, applying the transformation defined earlier, we obtain
\begin{eqnarray}
F_e(y)&=&f_e(y)e^{-g_e(y)},\label{fve}\\
f_e(y)&=& 1 - \frac{1-y}{2x_e}\mbox{erf}\left(\frac{2\mu_e x_e}{1-y}\right) + \frac{2\mu_e}{\sqrt{\pi}}e^{-\frac{4\mu_e^2 x_e^2}{(1-y)^2}},\quad
g_e(y)=\frac{\mu_e}{2}\left[1 -\mbox{erf}\left(\frac{2\mu_e x_e}{1-y}\right)+\frac{4\mu_e x_e}{\sqrt{\pi}(1-y)}e^{-\frac{4\mu_e^2 x_e^2}{(1-y)^2}}\right],\label{fv0e}\\
V_s(y) &=& \frac{(1-y)^2}{16}f_e(y)e^{-2g_e(y)}\left[(1-s^2)(1-y)\dot{f}_e(y)-(1-s)(1-y)f_e(y)\dot{g}_e(y)+\ell(\ell+1)\right].\label{fv1e}\\
\eta(y)&=&\frac{1+y}{1-y}+\frac{(1-y)e^{-\frac{\mu_e}{2}(x_e-1)}}{(\mu_e^2 x_e^2-1)(1+y)}.
\end{eqnarray}
\begin{table}
\caption{Characterization of the points $y = \pm 1$ for the functions involved in \eqref{S2oe}), \eqref{S1oe}, and \eqref{S0oe}. The abbreviations $z$ ord $n$ and $p$ ord $m$ denote a zero of order $n$ and a pole of order $m$, respectively.}
\begin{center}
\begin{tabular}{ | l | l | l | l |l |l | l | l}
\hline
$y$  & $F_e(y)$  & $V_s(y)$ & $\eta(y)$ & $S_{2e}(y)$ & $S_{1e}(y)$ & $S_{0e}(y)$\\ \hline
$-1$ & z \mbox{ord} 2 & z \mbox{ord} 2 & p \mbox{ord} 1 & z \mbox{ord} 6& z \mbox{ord} 4 & z \mbox{ord} 4 \\ \hline
$+1$ & $+1$  & z \mbox{ord} 2 & p \mbox{ord} 1 & $+1$ & p \mbox{ord} 2 & p \mbox{ord} 2\\ \hline
\end{tabular}
\label{table3}
\end{center}
\end{table}
Table~\ref{table3} indicates that the coefficients of the differential equation \eqref{ODEye} have a common zero of order 4 at $y = -1$, while $y = 1$ is a pole of order $2$ for the coefficients $S_{1e}(y)$ and $S_{0e}(y)$. Therefore, to apply the spectral method, it is necessary to multiply \eqref{ODEye} by $(1-y)^2/(1+y)^4$. This transformation leads to the following differential equation
\begin{equation}\label{ODEhynonee}
    M_{2e}(y)\ddot{\Phi}_{\Omega\ell s}(y)+M_{1e}(y)\dot{\Phi}_{\Omega\ell s}(y)+M_{0e}(y)\Phi_{\Omega\ell s}(y)=0,
\end{equation}
where
\begin{equation}\label{S210hononee}
  M_{2e}(y)=\frac{(1-y)^2}{4(1+y)^2}F_e^2(y),\quad
  M_{1e}(y)=i\Omega N_{1e}(y)+N_{0e}(y),\quad
  M_{0e}(y)=\Omega^2 C_{2e}(y)+i\Omega C_{1e}(y)+C_{0e}(y)
\end{equation}
with
\begin{eqnarray}
    N_{1e}(y)&=&\frac{1-y}{(1+y)^3}F_e^2(y)\left[x_e(1-y^2)\dot{\eta}(y)+2-a(1-y)\right],\quad
    N_{0e}(y)=\frac{F_e(y)}{4(1+y)^2}\frac{d}{dy}\left((1-y)^2 F_e(y)\right),\label{N0e}\\
    C_{2e}(y)&=&\frac{4x_e^2}{(1-y^2)^2}-\frac{F_e^2(y)}{(1+y)^4}\left[x_e(1-y^2)\dot{\eta}(y)+2-a(1-y)\right]^2,\label{C2e}\\
    C_{1e}(y)&=&\frac{x_e}{2}\left(\frac{1-y}{1+y}\right)^2 F_e(y)\dot{F}_e(y)\dot{\eta}(y)+\frac{x_e}{2}\frac{1-y}{(1+y)^2}F_e^2(y)\left[(1-y)\ddot{\eta}(y)-2\dot{\eta}(y)\right]+\nonumber\\
    &&\frac{1-y}{2(1+y)^3}\left[2-a(1-y)\right]F_e(y)\dot{F}_e(y)-\frac{F_e^2(y)}{(1+y)^4}\left[2-2a(1-y)+\frac{a}{2}(1-y)^2\right],\label{C1e}\\
    C_{0e}(y)&=&-\frac{4V_s(y)}{(1-y^2)^2}.\label{C0e}
\end{eqnarray}
It can be readily verified using Maple that
\begin{eqnarray}
    &&\lim_{y\to 1^{-}}M_{2e}(y)=0=\lim_{y\to -1^{+}}M_{2e}(y),\quad\lim_{y\to 1^{-}}M_{1e}(y)=\frac{1}{2}ix_e\Omega,\quad
    \lim_{y\to -1^{+}}M_{1e}(y)=i\Omega\Lambda_{1e},\\
    &&\lim_{y\to 1^{-}}M_{0e}(y)=A_{2e}\Omega^2 +A_{0e},\quad
     \lim_{y\to -1^{+}}M_{0e}(y)=B_{2e}\Omega^2+B_{0e},
\end{eqnarray}
where
\begin{eqnarray}
\Lambda_{1e}&=&-\frac{x_e}{2}\left(\mu_e^2 x_e^2-1\right)e^{\frac{\mu_e}{2}(x_e-1)},\quad
A_{2e}=\frac{1}{2}+\frac{x_e^2(4\mu_e^4 x_e^4-3\mu_e^3 x_e^3+2\mu_e^2 x_e^2+3\mu_e x_e-10)}{12\left(\mu_e^2 x_e^2-1\right)^2}e^{\frac{\mu_e}{2}(x_e-1)},\label{Acoefnonee}\\
A_{0e}&=&-\frac{\ell(\ell+1)}{16},\quad B_{0e}=-\frac{\ell(\ell+1)}{16}(\mu_e^2 x_e^2-1)e^{\mu_e(x_e-1)},\label{B0e}\\
B_{2e}&=&\frac{x^2_e\left(8\mu_e^8 x_e^8+12\mu_e^7 x_e^7+77\mu_e^6 x_e^6-96\mu_e^5 x_e^5-230\mu_e^4 x_e^4+180\mu_e^3 x_e^3+145\mu_e^2 x_e^2-96\mu_e x_e +32\right)}{144\left(\mu_e^2 x_e^2-1\right)^2}+\\
&&\frac{x_e(x_e+1)}{2}(\mu_e^2 x_e^2-1)e^{\frac{\mu_e}{2}(x_e-1)}.\label{Dcoefnonee}
\end{eqnarray}
Finally, to facilitate the application of the spectral method, we reformulate the differential equation \eqref{ODEhynonee} into the following form
\begin{equation}\label{TSCHe}
\widehat{L}^{(e)}_0\left[\Phi_{\Omega\ell s},\dot{\Phi}_{\Omega\ell s},\ddot{\Phi}_{\Omega\ell s}\right]+ i\widehat{L}^{(e)}_1\left[\Phi_{\Omega\ell s},\dot{\Phi}_{\Omega\ell s},\ddot{\Phi}_{\Omega\ell s}\right]\Omega+ \widehat{L}_2^{(e)}\left[\Phi_{\Omega\ell s},\dot{\Phi}_{\Omega\ell s},\ddot{\Phi}_{\Omega\ell s}\right]\Omega^2=0
\end{equation}
with
\begin{eqnarray}
\widehat{L}^{(e)}_0\left[\Phi_{\Omega\ell s},\dot{\Phi}_{\Omega\ell s},\ddot{\Phi}_{\Omega\ell s}\right]&=&\widehat{L}^{(e)}_{00}(y)\Phi_{\Omega\ell s}+\widehat{L}^{(e)}_{01}(y)\dot{\Phi}_{\Omega\ell s}+\widehat{L}^{(e)}_{02}(y)\ddot{\Phi}_{\Omega\ell s},\label{L0nonee}\\
\widehat{L}^{(e)}_1\left[\Phi_{\Omega\ell s},\dot{\Phi}_{\Omega\ell s},\ddot{\Phi}_{\Omega\ell s}\right]&=&\widehat{L}^{(e)}_{10}(y)\Phi_{\Omega\ell s}+\widehat{L}^{(e)}_{11}(y)\dot{\Phi}_{\Omega\ell s}+\widehat{L}^{(e)}_{12}(y)\ddot{\Phi}_{\Omega\ell s},\label{L1nonee}\\
\widehat{L}^{(e)}_2\left[\Phi_{\Omega\ell s},\dot{\Phi}_{\Omega\ell s},\ddot{\Phi}_{\Omega\ell s}\right]&=&\widehat{L}^{(e)}_{20}(y)\Phi_{\Omega\ell s}+\widehat{L}^{(e)}_{21}(y)\dot{\Phi}_{\Omega\ell s}+\widehat{L}^{(e)}_{22}(y)\ddot{\Phi}_{\Omega\ell s}.\label{L2nonee}
\end{eqnarray}
Furthermore, Table~\ref{table4} provides a summary of the $\widehat{L}^{(e)}_{ij}$ coefficients appearing in \eqref{L0nonee}-\eqref{L2nonee}, along with their limiting values at $y = \pm 1$.

\begin{table}
\caption{Definitions of the coefficients \(\widehat{L}^{(e)}_{ij}\) and their corresponding behaviour at the endpoints of the interval \(-1 \leq y \leq 1\). The symbols used here are defined in \eqref{S210hononee}-\eqref{B0e}.}
\begin{center}
\begin{tabular}{ | l | l | l | l |l |l | l | l}
\hline
$(i,j)$  & $\displaystyle{\lim_{y\to -1^+}}\widehat{L}^{(e)}_{ij}$  & $\widehat{L}^{(e)}_{ij}$ & $\displaystyle{\lim_{y\to 1^-}}\widehat{L}^{(e)}_{ij}$  \\ \hline
$(0,0)$ &  $B_{0e}$       & $C_{0e}$                  & $A_{0e}$\\ \hline
$(0,1)$ &  $0$            & $N_{0e}$                  & $0$\\ \hline
$(0,2)$ &  $0$            & $M_{2e}$                  & $0$\\ \hline 
$(1,0)$ &  $0$            & $C_{1e}$                  & $0$\\ \hline 
$(1,1)$ &  $\Lambda_{1e}$ & $N_{1e}$                  & $x_e/2$\\ \hline 
$(1,2)$ &  $0$            & $0$                       & $0$\\ \hline 
$(2,0)$ &  $B_{2e}$       & $C_{2e}$                  & $A_{2e}$\\ \hline
$(2,1)$ &  $0$            & $0$                       & $0$\\ \hline
$(2,2)$ &  $0$            & $0$                       & $0$\\ \hline
\end{tabular}
\label{table4}
\end{center}
\end{table}

\section{Numerical method}

To solve the differential eigenvalue problems \eqref{TSCH} and \eqref{TSCHe} and determine the QNMs and their corresponding frequencies $\Omega$, we discretize the differential operators $\widehat{L}_{j}[\cdot]$ and $\widehat{L}^{(e)}_{j}[\cdot]$ $j\in\{1, 2, 3\}$ as defined in equations (\ref{L0none})–(\ref{L2none}) and (\ref{L0nonee})–(\ref{L2nonee}), respectively. Since the problem is posed on the finite interval $[-1, 1]$ and only requires that the regular part of the QNM eigenfunction be regular at $y = \pm 1$, a Chebyshev-type spectral method \cite{Trefethen2000, Boyd2000} is a natural choice. Specifically, the function $\Phi_{\Omega \ell s}(y)$ is expanded in a truncated Chebyshev series
\begin{equation}
\Phi_{\Omega \ell s}(y) = \sum_{k=0}^{N} a_k T_k(y),
\end{equation} 
where $N \in \mathbb{N}$ is a numerical parameter, $\{a_k\}_{k=0}^{N} \subseteq \mathbb{R}$ are coefficients, and $\{T_k(y)\}_{k=0}^{N}$ are the Chebyshev polynomials of the first kind, defined by $T_k(y) = \cos(k \arccos y)$ for $y \in [-1, 1]$. Substituting this expansion into equations \eqref{TSCH} and \eqref{TSCHe} transforms the problem into an eigenvalue problem with polynomial coefficients. To reformulate it into a numerical framework, the collocation method \cite{Boyd2000} is employed. Instead of requiring the polynomial function in $y$ to vanish identically, this approach enforces vanishing at $N+1$ selected collocation points, which match the number of unknown coefficients $\{a_k\}_{k=0}^{N}$. For these collocation points, we use the Chebyshev roots grid \cite{Fox1968}  
\begin{equation}
y_k = -\cos\left(\frac{(2k+1)\pi}{2(N+1)}\right), \quad k \in \{0, 1, \ldots, N\}.
\end{equation}
An alternative option, also implemented in our codes, is the Chebyshev extrema grid
\begin{equation}
y_k = -\cos\left(\frac{k\pi}{N}\right), \quad k \in \{0, 1, \ldots, N\}.
\end{equation}  
Our numerical computations used the roots grid, though the theoretical performance of both choices is known to be comparable \cite{Fox1968, Boyd2000}. Applying the collocation method results in a classical matrix-based quadratic eigenvalue problem \cite{Tisseur2001}  
\begin{equation}\label{eq:eig}
(M_0 + iM_1\Omega + M_2\Omega^2)\mathbf{a} = \mathbf{0},
\end{equation}
where $M_j$ with $j\in\{0, 1, 2\}$ are square real matrices of size $(N+1) \times (N+1)$ that represent the spectral discretizations of \(\widehat{L}_{j}[\cdot]\) or \(\widehat{L}^{(e)}_{j}[\cdot]\). This eigenvalue problem is solved using the \texttt{polyeig} function in \textsc{Matlab}, yielding $2(N+1)$ potential eigenvalues for $\Omega$. To identify the physically meaningful QNM frequencies, root plots for different values of $N$ (e.g., $N = 100, 150, 200$) are overlapped, and consistent roots that remain stable across these values are selected.  To minimize numerical rounding errors and floating-point inaccuracies, all computations were performed using multiple precision arithmetic in \textsc{Maple}, interfaced with \textsc{Matlab} via the \textsc{Advanpix} toolbox \cite{mct2015}. The reported numerical results were calculated with a precision of 200 decimal digits.  

\section{Numerical results}

In this section, we present our numerical findings for the QNMs of non-extremal and extremal noncommutative geometry-inspired dirty black holes, computed using the spectral method described earlier, with a focus on understanding the effects of noncommutativity, horizon structure, and near-extremal configurations on the ringdown spectrum. Additionally, we compare these results with the QNMs of the noncommutative geometry-inspired Schwarzschild black hole to highlight the influence of the dirtiness in the spacetime geometry. The tables~\ref{table:0}-\ref{table:3b} provide several significant observations regarding the QNMs of noncommutative geometry-inspired dirty black holes across scalar, electromagnetic, and vector-type gravitational perturbations for nearly extremal and extremal cases. First of all, in the large mass parameter limit $\mu = 10^3$ (Schwarzschild limit), the QNM frequencies of noncommutative geometry-inspired dirty black holes converge closely to those of the classical Schwarzschild black hole, as shown in Table~\ref{table:0}. This result validates the spectral method employed and reinforces the consistency of the noncommutative model in approximating classical results. For moderate values of $\mu$, such as $\mu=1.95$ and $\mu=2.25$, deviations from the classical Schwarzschild and noncommutative Schwarzschild cases are evident. Tables~\ref{table:1}–\ref{table:1aa} show shifts in the frequencies' real and imaginary parts due to the dirtiness introduced by smeared energy distributions. As $\mu$ increases in the aforementioned range, the QNMs get closer to those of the noncommutative Schwarzschild black hole, indicating that the influence of additional matter distributions diminishes with increasing $\mu$. Moeover, Tables~\ref{table:1ext}–\ref{table:3b} signalize the appearance of purely imaginary frequencies $\Omega_R = 0$ in nearly extremal ($\mu=1.905$ and $\mu=1.91$) and extremal cases. These overdamped modes are characteristic of rapid decay without oscillations and already appear at low angular momentum quantum numbers $\ell$. They become especially pronounced for scalar perturbations in the extremal limit $\mu=\mu_e$ (see Table~\ref{table:2}), indicating significant modifications in the ringdown behaviour of the black hole.

For nearly extremal configurations $\mu\approx \mu_e$, Tables~\ref{table:1ext}–\ref{table:1aaext} reveal a gradual transition in the QNM spectrum. The presence of overdamped modes and their growth as $\mu$ approaches $\mu_e$ is indicative of the degenerate horizon structure. Although Table~\ref{table:3} does not display overdamped modes for $\ell=1$ and $\ell=2$ when using 200 Chebyshev polynomials, we found that such modes begin to appear for $\ell=1$ upon increasing the number of polynomials to 300. For example, in the case of $s=1=\ell$, $\mu=1.905$, and $N=300$, we identified a sequence of purely imaginary QNMs with $\Omega = -1.2986i, -1.4135i, -1.5284i, -1.6434i, -1.7576i$.  In the extremal case $\mu = \mu_e$, the QNM spectrum reveals some distinctive features. Overdamped modes appear with frequencies that differ from those of noncommutative Schwarzschild black holes. Moreover, the real and imaginary parts of the QNM frequencies show consistent deviations, reflecting the unique boundary conditions and geometry of extremal black hole configurations. Finally, we note that the overdamped modes identified in our analysis share similarities with the near-extremal frequencies reported by \cite{Cardoso2018PRL}. At the same time, these modes can be viewed as arising from the so-called zero-damped modes, which form a sequence of QNMs converging to purely imaginary values in the extremal limit \cite{Joykutty2022AHP}. Interestingly, such modes have also been observed in the nearly extremal regimes of various black hole spacetimes, including Reissner-Nordström \cite{Hod2010PLA, Hod2012PLB, Hod2015PLB}, Reissner-Nordström-de Sitter \cite{Cardoso2018PRD, Destounis2019PLB, Destounis2019JHEP, Joykutty2022AHP}, Kerr \cite{Hod2008PRDa, Hod2009PRD, Hod2011PRD, Yang2013PRDa, Yang2013PRDb}, Kerr-Newman \cite{Hod2008PLB, Dias2015PRL}, and black strings \cite{Wuthicharn2021IJMPD}.

Furthermore, the emergence of overdamped modes in the near-extremal and extremal regimes has important observational implications. In these regimes, the QNM spectrum is dominated by predominantly imaginary frequencies, indicating a rapid, non-oscillatory decay of perturbations. This behavior deviates significantly from the standard damped-oscillatory ringdown expected for classical black holes. Recent studies (see, e.g. \cite{Banerjee2025PDU,Kain2021PRD}) have suggested that such departures from the canonical signal could serve as potential indicators of quantum gravitational corrections or modified horizon structures. In our model, the overdamped modes arise due to noncommutative corrections that modify the effective potential, particularly as the black hole approaches extremality. Thus, a careful analysis of the ringdown phase in gravitational wave observations may provide a novel observational window into the effects of noncommutative geometry, offering valuable insights into the underlying quantum structure of spacetime.

We also observe that the effect of dirtiness becomes particularly prominent in the extremal regime. While the fundamental modes remain nearly identical between the extreme dirty case and the corresponding extreme noncommutative geometry-inspired Schwarzschild black hole—with deviations showing up mainly in higher overtones—the overdamped modes exhibit clear differences in the extreme configuration. In contrast, in the nearly-extremal regime (e.g., \(\mu = 1.905\) as shown in Tables IX, X, and XI), the overdamped modes of the nearly extreme dirty black hole closely match those of the nearly extreme noncommutative geometry-inspired case. This indicates that the dirtiness significantly impacts the spectrum only when the black hole reaches the extremal limit.

We observe that scalar, electromagnetic, and gravitational perturbations exhibit distinct QNM spectra, with differences in both the oscillatory component $\Omega_R$ and the damping rate $\Omega_I$ that are consistent across nearly extremal and extremal cases (see Tables~\ref{table:1aext}–\ref{table:3b}), indicating that the impact of noncommutativity is not limited to a single perturbation type. In particular, as the black hole approaches extremality, the event and Cauchy horizons converge into a degenerate horizon. This degeneracy alters the effective potential by reducing the restoring force that typically induces oscillatory behavior, resulting in a QNM spectrum dominated by rapidly decaying, purely imaginary modes. These overdamped modes arise from the combined effects of horizon degeneracy and noncommutative corrections, in good agreement with previous studies \cite{Leung2011PRD,Kain2021PRD,Banerjee2025PDU,Banerjee2025CQG,Tangphati2024PDU, GleasonPRD2022}. Last but not least, the spectral method shows excellent agreement with other established techniques (e.g., continued fraction and WKB methods) in the large-\(\mu\) limit, as seen in Table~\ref{table:0}, and is capable of capturing subtle variations in the QNM spectrum for moderate and small \(\mu\), confirming its robustness and precision. In summary, the distinctive overdamped modes and shifts in the QNM spectra induced by noncommutativity and the dirtiness of the black hole may serve as valuable observational signatures in gravitational-wave data, with the behavior near extremality offering a promising diagnostic tool for differentiating noncommutative black holes from their classical counterparts.

\begin{table}
\centering
\caption{This table presents a comparison of the quasinormal frequencies for scalar perturbations (\(s = 0\)) of the classic Schwarzschild black hole and a noncommutative geometry-inspired dirty black hole with large mass parameter $\mu = 10^3$ an event horizon at $x_h = 1$, calculated using the spectral method with $200$ polynomials and accuracy of $200$ digits (see last column). The third column shows the numerical values computed by \cite{Leaver1985PRSLA} using the continued fraction method. The fourth column includes third-order WKB results from \cite{IYER1987PRD}, while the fifth column provides numerical values from \cite{Mamani2022EPJC} obtained via the spectral method with 40 polynomials. The sixth column lists the quasinormal modes for the noncommutative geometry-inspired Schwarzschild black hole as reported by \cite{Batic2024EPJC}. The notation "N/A" signifies unavailable data, and the subscripts \(S\) and \(NCS\) denote the Schwarzschild case and the noncommutative geometry-inspired Schwarzschild black hole, respectively.}
\label{table:0}
 \vspace*{1em}
 \begin{tabular}{||c|c|c|c|c|c|c||} 
 \hline
 $\ell$ & $n$ & $\Omega_S$ \cite{Leaver1985PRSLA} & $\Omega_S$ \cite{IYER1987PRD} & $\Omega_S$ \cite{Mamani2022EPJC} & $\Omega_{NCS}$, $\mu=10^3$ \cite{Batic2024EPJC}& $\Omega$, $\mu=10^3$\\ [0.5ex] 
 \hline\hline
 \rule{0pt}{3ex} 
 $0$ & $0$ & $0.1105-0.1049i$ & $0.1046-0.1152i$ & $0.1105-0.1049i$ & $0.1104549-0.1048957i$ & $0.110455-0.10490i$ \\ 
     & $1$ & $0.0861-0.3481i$ & $0.0892-0.3550i$ & \mbox{N/A}       & $0.0861169-0.3480525i$ & $0.086137-0.34804i$ \\
 $1$ & $0$ & $0.2929-0.0977i$ & $0.2911-0.0980i$ & $0.2929-0.0977i$ & $0.2929361-0.0976599i$ & $0.292936-0.09766i$ \\
     & $1$ & $0.2645-0.3063i$ & $0.2622-0.3704i$ & $0.2645-0.3063i$ & $0.2644487-0.3062574i$ & $0.264449-0.30626i$ \\
     & $2$ & $0.2295-0.5401i$ & $0.2235-0.5268i$ & \mbox{N/A}       & $0.2295393-0.5401334i$ & $0.229539-0.54013i$ \\ 
     & $3$ & $0.2033-0.7883i$ & $0.1737-0.7486i$ & \mbox{N/A}       & $0.2032584-0.7882978i$ & $0.203259-0.78830i$ \\
 $2$ & $0$ & $0.4836-0.0968i$ & $0.4832-0.0968i$ & $0.4836-0.0968i$ & $0.4836439-0.0967588i$ & $0.483644-0.09676i$ \\
     & $1$ & $0.4639-0.2956i$ & $0.4632-0.2958i$ & $0.4639-0.2956i$ & $0.4638506-0.2956039i$ & $0.463851-0.29560i$ \\
     & $2$ & $0.4305-0.5086i$ & $0.4317-0.5034i$ & $0.4305-0.5086i$ & $0.4305441-0.5085584i$ & $0.430544-0.50856i$ \\
     & $3$ & $0.3939-0.7381i$ & $0.3926-0.7159i$ & \mbox{N/A}       & $0.3938631-0.7380966i$ & $0.393863-0.73810i$ \\ [0.5ex] 
 \hline
 \end{tabular}
\end{table}

\begin{table}
\centering
\caption{QNMs for scalar perturbations (\(s = 0\)) of the noncommutative geometry-inspired dirty black hole metric (nonextreme case) are presented for various values of the mass parameter \(\mu\). The numerical values in the third and fifth columns are computed using the spectral method with $200$ polynomials and an accuracy of $200$ digits. For comparison, the fourth and sixth columns provide the corresponding results for the noncommutative geometry-inspired Schwarzschild black hole \cite{Batic2024EPJC}. In this context, \(\Omega\) denotes the dimensionless frequency as defined in equation (\ref{ourODE}). The subscript \(NCS\) indicates that the quantity pertains to the noncommutative geometry-inspired Schwarzschild black hole. Note that the mass parameter values \(\mu = 1.95\) and \(\mu = 2.25\) correspond to event horizons located at \(x_h = 0.88869\) and \(x_h = 0.97868\), respectively.}
\label{table:1}
\vspace*{1em}
\begin{tabular}{||c|c|c|c|c|c||} 
\hline
$\ell$ & 
$n$    & 
$\Omega$, $\mu=1.95$  & 
$\Omega_{NCS}$, $\mu=1.95$ \cite{Batic2024EPJC} &
$\Omega$, $\mu=2.25$  &
$\Omega_{NCS}$, $\mu=2.25$ \cite{Batic2024EPJC}\\ [0.5ex] 
\hline\hline
\rule{0pt}{3ex} 
$0$ & $0$ & $0.1033-0.0938i$ & $0.1041-0.0930i$ & $0.1098-0.1008i$ & $0.1101-0.1010i$ \\ 
    & $1$ & $0.0228-0.3505i$ & $0.0343-0.3374i$ & $0.0503-0.3298i$ & $0.0595-0.3298i$ \\
$1$ & $0$ & $0.2879-0.0927i$ & $0.2890-0.0908i$ & $0.2912-0.0962i$ & $0.2916-0.0958i$ \\
    & $1$ & $0.2343-0.2882i$ & $0.2381-0.2807i$ & $0.2530-0.2989i$ & $0.2556-0.2971i$ \\
    & $2$ & $0.1429-0.5247i$ & $0.1480-0.5047i$ & $0.1842-0.4272i$ & $0.1961-0.5186i$ \\ 
    & $3$ & $0.0608-0.8040i$ & $0.0611-0.7768i$ & $0.1241-0.8093i$ & $0.1056-0.7762i$ \\
$2$ & $0$ & $0.4797-0.0937i$ & $0.4806-0.0916i$ & $0.4823-0.0962i$ & $0.4825-0.0957i$ \\
    & $1$ & $0.4451-0.2840i$ & $0.4477-0.2667i$ & $0.4571-0.2930i$ & $0.4581-0.2909i$ \\
    & $2$ & $0.3778-0.4868i$ & $0.3833-0.4607i$ & $0.4095-0.5018i$ & $0.4134-0.4965i$ \\
    & $3$ & $0.2855-0.7181i$ & $0.2925-0.6865i$ & $0.3435-0.7296i$ & $0.3537-0.7160i$ \\ [1ex]
 \hline
 \end{tabular}
\end{table}

\begin{table}
\centering
\caption{QNMs for electromagnetic perturbations ($s = 1$) of the noncommutative geometry-inspired dirty black hole metric (nonextreme case) are presented for various values of the mass parameter \(\mu\). The numerical values in the third and fifth columns are computed using the spectral method with $200$ polynomials and an accuracy of $200$ digits. For comparison, the fourth and sixth columns provide the corresponding results for the noncommutative geometry-inspired Schwarzschild black hole \cite{Batic2024EPJC}. Here, \(\Omega\) denotes the dimensionless frequency as defined in equation (\ref{ourODE}). The subscript \(NCS\) refers to the noncommutative geometry-inspired Schwarzschild black hole. Notice that the mass parameter values $\mu = 1.95$ and $\mu = 2.25$ correspond to event horizons located at $x_h = 0.88869$ and $x_h = 0.97868$, respectively.} 
\label{table:1a}
\vspace*{1em}
\begin{tabular}{||c|c|c|c|c|c||}  
\hline
$\ell$ & 
$n$ & 
$\Omega$, $\mu=1.95$ & 
$\Omega_{NCS}$, $\mu=1.95$ \cite{Batic2024EPJC} &
$\Omega$, $\mu=2.25$  &
$\Omega_{NCS}$, $\mu=2.25$ \cite{Batic2024EPJC}\\ [0.5ex] 
\hline\hline
\rule{0pt}{3ex} 
$1$ & $0$ & $0.2453-0.0849i$ & $0.2469-0.0831i$ & $0.2473-0.0902i$ & $0.2478-0.0899i$ \\
    & $1$ & $0.1883-0.2646i$ & $0.1945-0.2569i$ & $0.2059-0.2820i$ & $0.2090-0.2810i$ \\
    & $2$ & $0.0858-0.4894i$ & $0.0965-0.4661i$ & $0.1346-0.4970i$ & $0.1491-0.4936i$ \\
    & $3$ & $0.0215-0.7792i$ & $0.0181-0.6972i$ & $0.0650-0.7914i$ & $0.0656-0.7133i$ \\
$2$ & $0$ & $0.4541-0.1091i$ & $0.4552-0.0887i$ & $0.4564-0.0942i$ & $0.4566-0.0936i$ \\
    & $1$ & $0.4188-0.2758i$ & $0.4220-0.2682i$ & $0.4301-0.2868i$ & $0.4314-0.2848i$ \\
    & $2$ & $0.3494-0.4729i$ & $0.3566-0.4561i$ & $0.3809-0.4916i$ & $0.3855-0.4866i$ \\
    & $3$ & $0.2531-0.6995i$ & $0.2637-0.6656i$ & $0.3126-0.7150i$ & $0.3254-0.7022i$ \\ [1ex]
 \hline
 \end{tabular}
\end{table}

\begin{table}
\centering
\caption{QNMs for vector-type gravitational perturbations ($s = 2$) of the noncommutative geometry-inspired dirty black hole metric (nonextreme case) are presented for various values of the mass parameter \(\mu\). The numerical values in the third and fifth columns are computed using the spectral method with $200$ polynomials and an accuracy of $200$ digits. For comparison, the fourth and sixth columns provide the corresponding results for the noncommutative geometry-inspired Schwarzschild black hole \cite{Batic2024EPJC}. Here, \(\Omega\) denotes the dimensionless frequency as defined in equation (\ref{ourODE}). The subscript \(NCS\) refers to the noncommutative geometry-inspired Schwarzschild black hole. Notice that the mass parameter values $\mu = 1.95$ and $\mu = 2.25$ correspond to event horizons located at $x_h = 0.88869$ and $x_h = 0.97868$, respectively.} 
\label{table:1aa}
\vspace*{1em}
\begin{tabular}{||c|c|c|c|c|c||}  
\hline
$\ell$ & 
$n$ & 
$\Omega$, $\mu = 1.95$  & 
$\Omega_{NCS}$, $\mu = 1.95$ \cite{Batic2024EPJC} &
$\Omega$, $\mu = 2.25$  &
$\Omega_{NCS}$, $\mu = 2.25$ \cite{Batic2024EPJC} \\ [0.5ex] 
\hline\hline
\rule{0pt}{3ex} 
$2$ & $0$  & $0.3714-0.0801i$ &  $0.3723-0.0776i$ & $0.3725-0.0867i$ & $0.3726-0.0861i$ \\
    & $1$  & $0.3358-0.2405i$ &  $0.3411-0.2316i$ & $0.3421-0.2631i$ & $0.3437-0.2611i$ \\
    & $2$  & $0.2701-0.4105i$ &  $0.2862-0.3945i$ & $0.2903-0.4497i$ & $0.2973-0.4462i$ \\
    & $3$  & $0.1816-0.6172i$ &  $0.2073-0.5902i$ & $0.2291-0.6530i$ & $0.2477-0.6492i$ \\
$3$ & $0$  & $0.5972-0.0880i$ &  $0.5979-0.0857i$ & $0.5985-0.0918i$ & $0.5985-0.0913i$ \\
    & $1$  & $0.5707-0.2650i$ &  $0.5730-0.2570i$ & $0.5781-0.2775i$ & $0.5786-0.2755i$ \\
    & $2$  & $0.5188-0.4460i$ &  $0.5252-0.4298i$ & $0.5392-0.4697i$ & $0.5417-0.4649i$ \\
    & $3$  & $0.4426-0.6388i$ &  $0.4569-0.6104i$ & $0.4846-0.6720i$ & $0.4923-0.6632i$ \\
    & $4$  & $0.3464-0.8541i$ &  $0.3693-0.8090i$ & $0.4169-0.8879i$ & $0.4347-0.8720i$ \\ [1ex] 
 \hline
 \end{tabular}
\end{table}

\begin{table}
\centering
\caption{QNMs for scalar perturbations ($s = 0$) of the noncommutative geometry-inspired dirty black hole metric in the nearly extremal case are presented for various values of the mass parameter $\mu$. These numerical values were computed using the spectral method with $200$ polynomials and an accuracy of $200$ digits. For comparison, the fourth and sixth columns provide the corresponding results for the noncommutative geometry-inspired Schwarzschild black hole. Here, $\Omega$ represents the dimensionless frequency as defined in equation \eqref{ourODE}, and the subscript $NCS$ indicates that the quantity refers to the noncommutative geometry-inspired Schwarzschild black hole. Additionally, the mass parameter values $\mu = 1.91$ and $\mu = 1.905$ correspond to event horizon locations at $x_h = 0.83098$ and $x_h = 0.80849$, respectively.}
\label{table:1ext}
\vspace*{1em}
\begin{tabular}{||c|c|c|c|c|c||} 
\hline
$\ell$ & 
$n$    & 
$\Omega$, $\mu = 1.91$  & 
$\Omega_{NCS}$, $\mu = 1.91$ &
$\Omega$, $\mu = 1.905$  &
$\Omega_{NCS}$, $\mu = 1.905$ \\ [0.5ex] 
\hline\hline
\rule{0pt}{3ex} 
$0$ & $0$ & $0.1022-0.0936i$ & $0.1029-0.0928i$ & $0.1021-0.0936i$ & $0.1028-0.0928i$ \\
    & $1$ & $0.0000-0.5789i$ & $0.0000-0.5789i$ & $0.0000-0.3148i$ & $0.0000-0.3148i$ \\
$1$ & $0$ & $0.2871-0.0919i$ & $0.2883-0.0897i$ & $0.2870-0.0918i$ & $0.2882-0.0896i$ \\
    & $1$ & $0.2310-0.2865i$ & $0.2346-0.2781i$ & $0.2306-0.2863i$ & $0.2342-0.2778i$ \\
    & $2$ & $0.0000-0.3645i$ & $0.1410-0.5024i$ & $0.0000-0.3645i$ & $0.0000-0.3645i$ \\ 
    & $3$ & $0.0000-0.6548i$ & $0.0000-0.6548i$ & $0.0000-0.6548i$ & $0.0000-0.6548i$ \\
$2$ & $0$ & $0.4792-0.0930i$ & $0.4802-0.0905i$ & $0.4791-0.0929i$ & $0.4802-0.0903i$ \\
    & $1$ & $0.4426-0.2820i$ & $0.4451-0.2736i$ & $0.4423-0.2818i$ & $0.4448-0.2732i$ \\
    & $2$ & $0.3720-0.4841i$ & $0.3770-0.4663i$ & $0.0000-0.3945i$ & $0.0000-0.3945i$ \\
    & $3$ & $0.0000-0.7078i$ & $0.0000-0.7076i$ & $0.0000-0.7078i$ & $0.0000-0.7078i$ \\ [1ex]
 \hline
 \end{tabular}
\end{table}

\begin{table}
\centering
\caption{QNMs for electromagnetic perturbations ($s = 1$) of the noncommutative geometry-inspired dirty black hole metric in the nearly extremal case are presented for various mass parameter $\mu$ values. These numerical values were computed using the spectral method with $200$ polynomials and an accuracy of $200$ digits. For comparison, the fourth and sixth columns provide the corresponding results for the noncommutative geometry-inspired Schwarzschild black hole. Here, $\Omega$ represents the dimensionless frequency as defined in equation \eqref{ourODE}, and the subscript $NCS$ indicates that the quantity refers to the noncommutative geometry-inspired Schwarzschild black hole. Additionally, the mass parameter values $\mu = 1.91$ and $\mu = 1.905$ correspond to event horizon locations at $x_h = 0.83098$ and $x_h = 0.80849$, respectively.} 
\label{table:1aext}
\vspace*{1em}
\begin{tabular}{||c|c|c|c|c|c||}  
\hline
$\ell$ & 
$n$    & 
$\Omega$, $\mu = 1.91$ & 
$\Omega_{NCS}$, $\mu = 1.91$ &
$\Omega$, $\mu = 1.905$ &
$\Omega_{NCS}$, $\mu = 1.905$ \\ [0.5ex] 
\hline\hline
\rule{0pt}{3ex} 
$1$ & $0$ & $0.2448-0.0838i$ & $0.2465-0.0815i$ & $0.2447-0.0837i$ & $0.2465-0.0813i$ \\
    & $1$ & $0.1851-0.2621i$ & $0.1909-0.2533i$ & $0.0000-0.3517i$ & $0.1905-0.2528i$ \\
    & $2$ & $0.0000-0.6326i$ & $0.0895-0.4633i$ & $0.0000-0.6326i$ & $0.0000-0.3517i$ \\
    & $3$ & $0.0000-0.9005i$ & $0.0000-0.6326i$ & $0.0000-0.8989i$ & $0.0000-0.6326i$ \\
$2$ & $0$ & $0.4537-0.0901i$ & $0.4550-0.0875i$ & $0.4536-0.0900i$ & $0.4549-0.0873i$ \\
    & $1$ & $0.4164-0.2734i$ & $0.4196-0.2646i$ & $0.4160-0.2731i$ & $0.4192-0.2641i$ \\
    & $2$ & $0.3435-0.4697i$ & $0.3503-0.4510i$ & $0.0000-0.3904i$ & $0.0000-0.3904i$ \\
    & $3$ & $0.0000-0.6997i$ & $0.0000-0.6995i$ & $0.0000-0.6997i$ & $0.0000-0.6997i$ \\ [1ex]
 \hline
 \end{tabular}
\end{table}

\begin{table}
\centering
\caption{QNMs for vector-type gravitational perturbations ($s = 2$) of the noncommutative geometry-inspired dirty black hole metric in the nearly extremal case are presented for various values of the mass parameter $\mu$. These numerical values were computed using the spectral method with $200$ polynomials and an accuracy of $200$ digits. For comparison, the fourth and sixth columns provide the corresponding results for the noncommutative geometry-inspired Schwarzschild black hole. Here, $\Omega$ represents the dimensionless frequency as defined in equation \eqref{ourODE}, and the subscript $NCS$ indicates that the quantity refers to the noncommutative geometry-inspired Schwarzschild black hole. Additionally, the mass parameter values $\mu = 1.91$ and $\mu = 1.905$ correspond to event horizon locations at $x_h = 0.83098$ and $x_h = 0.80849$, respectively.} 
\label{table:1aaext}
\vspace*{1em}
\begin{tabular}{||c|c|c|c|c|c||}  
\hline
$\ell$ & 
$n$    & 
$\Omega$, $\mu = 1.91$  & 
$\Omega_{NCS}$, $\mu = 1.91$ &
$\Omega$, $\mu = 1.905$ &
$\Omega_{NCS}$, $\mu = 1.905$ \\ [0.5ex] 
\hline\hline
\rule{0pt}{3ex} 
$2$ & $0$ & $0.3712-0.0785i$ & $0.3723-0.0753i$ & $0.3712-0.0783i$ & $0.3723-0.0750i$ \\
    & $1$ & $0.3341-0.2359i$ & $0.3396-0.2250i$ & $0.3339-0.2353i$ & $0.3394-0.2242i$ \\
    & $2$ & $0.2658-0.4047i$ & $0.2818-0.3865i$ & $0.0000-0.3762i$ & $0.0000-0.3762i$ \\
    & $3$ & $0.0000-0.6731i$ & $0.0000-0.6731i$ & $0.0000-0.6731i$ & $0.0000-0.6731i$ \\
$3$ & $0$ & $0.5970-0.0870i$ & $0.5979-0.0841i$ & $0.5970-0.0869i$ & $0.5979-0.0839i$ \\
    & $1$ & $0.5691-0.2620i$ & $0.5714-0.2524i$ & $0.5688-0.2616i$ & $0.5712-0.2517i$ \\
    & $2$ & $0.5143-0.4414i$ & $0.5206-0.4225i$ & $0.5138-0.4409i$ & $0.5200-0.4216i$ \\
    & $3$ & $0.4345-0.6333i$ & $0.4484-0.6017i$ & $0.0000-0.4055i$ & $0.0000-0.4055i$ \\
    & $4$ & $0.0000-0.7291i$ & $0.0000-0.7288i$ & $0.0000-0.7291i$ & $0.0000-0.7290i$ \\ [1ex]
 \hline
 \end{tabular}
\end{table}


\begin{table}
\centering
\caption{This table compares the quasinormal frequencies for scalar perturbations ($s = 0$) of the extremal noncommutative geometry-inspired dirty black hole (third column) with those of the noncommutative geometry-inspired Schwarzschild black hole (last column) as reported in \cite{Batic2024EPJC}. These values were computed using the spectral method with $200$ polynomials and an accuracy of $200$ digits. The subscript $NCS$ refers to the noncommutative geometry-inspired Schwarzschild black hole. Here, $\Omega$ denotes the dimensionless frequency as defined in equation \eqref{ourODEe}. In the extreme case, the mass parameter is $\mu_e = 1.904119076\ldots$ while the horizon is located at $x_e= 0.7936575898\ldots$.}
\label{table:2}
 \vspace*{1em}
 \begin{tabular}{||c|c|c|c||} 
 \hline
 $\ell$ & $n$ & $\Omega$  & $\Omega_{NCS}$ \cite{Batic2024EPJC} \\ [0.5ex] 
 \hline\hline
 \rule{0pt}{3ex} 
$0$ & $0$ & $0.1020-0.0935i$  & $0.1027 - 0.0928i$ \\ 
    & $1$ & $0.0000-0.7182i$  & $0.0299 - 0.3374i$ \\
$1$ & $0$ & $0.2870-0.0918i$  & $0.2882 - 0.0895i$ \\
    & $1$ & $0.2306-0.2862i$  & $0.2341 - 0.2777i$ \\
    & $2$ & $0.1354-0.5231i$  & $0.1400 - 0.5021i$ \\ 
    & $3$ & $0.0000-0.9842i$  & $0.0317 - 0.5720i$ \\
$2$ & $0$ & $0.4791-0.0929i$  & $0.4802 - 0.0903i$ \\
    & $1$ & $0.4422-0.2817i$  & $0.4447 - 0.2731i$ \\
    & $2$ & $0.3711-0.4837i$  & $0.3760 - 0.4657i$ \\
    & $3$ & $0.2747-0.7155i$  & $0.0000 - 0.6732i$ \\
    & $4$ & $0.1732-0.9862i$  & $0.2806 - 0.6818i$ \\ [0.5ex]
 \hline
 \end{tabular}
\end{table}


\begin{table}
\centering
\caption{This table presents the quasinormal frequencies for electromagnetic perturbations ($s = 1$) of the extremal noncommutative geometry-inspired dirty black hole (third column) with those of the noncommutative geometry-inspired Schwarzschild black hole (last column) as reported in \cite{Batic2024EPJC}. These values were computed using the spectral method with $200$ polynomials and an accuracy of $200$ digits. The subscript $NCS$ refers to the noncommutative geometry-inspired Schwarzschild black hole. Here, $\Omega$ denotes the dimensionless frequency as defined in equation \eqref{ourODEe}. In the extreme case, the mass parameter is $\mu_e = 1.904119076\ldots$ while the horizon is located at $x_e= 0.7936575898\ldots$.}
\label{table:3}
 \vspace*{1em}
 \begin{tabular}{||c|c|c|c||} 
 \hline
 $\ell$ & $n$ & $\Omega$ & $\Omega_{NCS}$ \cite{Batic2024EPJC} \\ [0.5ex] 
 \hline\hline
 \rule{0pt}{3ex} 
$1$ & $0$ & $0.2447-0.0836i$ & $0.2465-0.0813i$ \\ 
    & $1$ & $0.1846-0.2617i$ & $0.1904-0.2527i$ \\
    & $2$ & $0.0783-0.4871i$ & $0.0883-0.4627i$ \\
$2$ & $0$ & $0.4536-0.0900i$ & $0.4549-0.0872i$ \\
    & $1$ & $0.4160-0.2731i$ & $0.4192-0.2640i$ \\
    & $2$ & $0.3426-0.4692i$ & $0.3493-0.4503i$ \\ 
    & $3$ & $0.2420-0.6963i$ & $0.0000-0.6456i$ \\
    & $4$ & $0.1415-0.9578i$ & $0.2516-0.6601i$ \\ [0.5ex] 
 \hline
 \end{tabular}
\end{table}


\begin{table}
\centering
\caption{This table presents the quasinormal frequencies for vector-type gravitational perturbations ($s = 2$) of the extremal noncommutative geometry-inspired dirty black hole (third column) with those of the noncommutative geometry-inspired Schwarzschild black hole (last column) as reported in \cite{Batic2024EPJC}. These values were computed using the spectral method with $200$ polynomials and an accuracy of $200$ digits. The subscript $NCS$ refers to the noncommutative geometry-inspired Schwarzschild black hole. Here, $\Omega$ denotes the dimensionless frequency as defined in equation \eqref{ourODEe}. In the extreme case, the mass parameter is $\mu_e = 1.904119076\ldots$ while the horizon is located at $x_e= 0.7936575898\ldots$.}
\label{table:3b}
 \vspace*{1em}
 \begin{tabular}{||c|c|c|c||} 
 \hline
 $\ell$ & $n$ & $\Omega$ & $\Omega_{NSC}$ \cite{Batic2024EPJC} \\ [0.5ex] 
 \hline\hline
 \rule{0pt}{3ex} 
$2$ & $0$ & $0.3712-0.0782i$ & $0.3723-0.0749i$ \\ 
    & $1$ & $0.3338-0.2352i$ & $0.3394-0.2240i$ \\
    & $2$ & $0.2651-0.4039i$ & $0.2811-0.3853i$ \\
    & $3$ & $0.1731-0.6123i$ & $0.1975-0.5832i$ \\
$3$ & $0$ & $0.5970-0.0869i$ & $0.5979-0.0839i$ \\
    & $1$ & $0.5688-0.2615i$ & $0.5711-0.2516i$ \\
    & $2$ & $0.5136-0.4406i$ & $0.5199-0.4214i$ \\ 
    & $3$ & $0.4333-0.6325i$ & $0.4471-0.6005i$ \\
    & $4$ & $0.3322-0.8490i$ & $0.3546-0.7997i$ \\ [0.5ex] 
 \hline
 \end{tabular}
\end{table}

\section{Conclusions and outlook}

Our study explored the QNMs of noncommutative geometry-inspired dirty black holes, which provided insights into their spectral properties. By means of a high-precision spectral method based on Chebyshev polynomials, we analyzed scalar, electromagnetic, and gravitational perturbations in both non-extremal and extremal configurations. We identified purely imaginary overdamped modes, particularly pronounced in near-extremal and extremal regimes. These modes signalize rapid decay without oscillations and are indicative of distinct geometric and boundary conditions imposed by noncommutative effects and the smeared matter distributions. Such features highlight the deviations from classical Schwarzschild black holes. For large mass parameters ($\mu = 10^3$), the QNMs converge to the classical Schwarzschild frequencies, demonstrating the robustness and accuracy of the numerical method we employed. This validates the noncommutative model as a consistent extension of classical black hole solutions. The extremal and nearly extremal cases present unique spectral features, including significant shifts in QNM frequencies compared to noncommutative Schwarzschild black holes. These results suggest that the behaviour near extremality could provide observational signatures for distinguishing noncommutative black holes in gravitational wave data. Finally, our findings show how the spectral method is capable of capturing fine details of QNM spectra across various perturbation types. It proves particularly effective in detecting small variations and overdamped modes, which might be challenging for other numerical techniques.

The results of this work open several avenues for future research. First of all, with the advancement of gravitational-wave astronomy, exploring the detectability of the distinctive QNMs associated with noncommutative black holes could provide a practical test of these theoretical models. Detailed waveform analysis and comparison with experimental data will be crucial.  Furthermore, the effectiveness of the spectral method, which has been shown not only in this study but also in \cite{Mamani2022EPJC, Batic2024CGG, Batic2024EPJC, Batic2024PRD} in connection with other spacetimes, highlights its potential for broader applications. These include exploring higher-dimensional spacetimes, rotating black holes, and black holes with other quantum-inspired modifications. Moreover, extending the analysis to include nonlinear perturbations could offer deeper insights into the stability and dynamics of these black holes under more general conditions. Last but not least, further investigations into the role of noncommutativity in resolving singularities and its interplay with other quantum gravity approaches could refine our understanding of black hole physics. We conclude by emphasizing that our work deepens the understanding of QNMs in noncommutative geometry-inspired black holes and provides a robust computational framework applicable to a wide range of problems in modified theories of gravity.

\section*{Code availability}

All analytical computations reported in this manuscript have been rechecked in the computer algebra system \textsc{Maple}. Two \textsc{Maple} sheets corresponding to the extreme and non-extreme cases can be found in supplementary materials \emph{and} in the repository below. The discretization of differential operators \eqref{L0none}-\eqref{L2none} and \eqref{L0nonee}-\eqref{L2nonee} using the Tchebyshev-type spectral method is equally performed in \textsc{Maple} computer algebra system. Finally, the numerical solution of the resulting quadratic eigenvalue problem \eqref{eq:eig} is performed in \textsc{Matlab} software using the \texttt{polyeig} function. All these materials are freely accessible at the following \texttt{GitHub} repository:

\begin{itemize}
    \item \url{https://github.com/dutykh/dirtybh/}
\end{itemize}

\end{document}